%% file: bbatnlo.tex
\documentclass{elsart}
\usepackage{latexsym}
\usepackage{amssymb}
\usepackage{graphicx}
\usepackage{epsfig}

\newcommand{\beq}[0]{\begin{equation}}
\newcommand{\eeq}[0]{\end{equation}}

\newcommand{\ourcal}{\mathcal}

\newcommand{\oa}{${\ourcal O}(\alpha)$}
\newcommand{\oatwo}{${\ourcal O}(\alpha^2)$}
\newcommand{\be}{\begin{equation}}
\newcommand{\ee}{\end{equation}}
\newcommand{\bea}{\begin{eqnarray}}
\newcommand{\eea}{\end{eqnarray}}
\newcommand{\myref}[1]{(\ref{#1})}
\newcommand{\BABAYAGA}{$\tt BABAYAGA$}
\newcommand{\BHWIDE}{$\tt BHWIDE$}
\newcommand{\LABSPV}{$\tt LABSPV$}
\begin{document}
\flushright{FNT/T-2006/05}
\begin{frontmatter}
\title{Matching perturbative and Parton Shower corrections to Bhabha
process at flavour factories}
\author[1]{Giovanni Balossini}
\author[2,1]{Carlo M. Carloni Calame\corauthref{me}}
\author[1,2]{Guido Montagna}
\author[2,1]{Oreste Nicrosini}
\author[2,1]{Fulvio Piccinini}
\address[1]{Dipartimento di Fisica Nucleare e Teorica, Universit\`a di Pavia\\
Via A. Bassi 6, 27100, Pavia, Italy}
\address[2]{Istituto Nazionale di Fisica Nucleare, sezione di Pavia\\
Via A. Bassi 6, 27100, Pavia, Italy}
\corauth[me]{Corresponding author. E-mail address:
$\tt carlo.carloni.calame@pv.infn.it$}
\begin{abstract}
We report on a high-precision calculation of the Bhabha process in Quantum
Electrodynamics, of interest for precise luminosity
determination of electron-positron colliders
involved in $R$ measurements in the region of hadronic
resonances.
The calculation is based on the matching of exact next-to-leading 
order corrections with a Parton Shower algorithm. The accuracy of the approach
is demonstrated in comparison with existing independent calculations 
and through a detailed analysis of the main components of theoretical
uncertainty, including two-loop corrections, hadronic vacuum
polarization and light pair contributions. The calculation is
implemented in an improved version of the event generator $\tt BABAYAGA$
with a theoretical accuracy of the order of 0.1\%.
The generator is now available for
high-precision simulations of the Bhabha process at flavour factories.
\end{abstract}
\begin{keyword}
Quantum Electrodynamics; 
Bhabha process; luminosity; next-to-leading
order corrections; two-loop corrections; Parton 
Shower\\
{\sc pacs}: 12.15.Lk, 12.20.-m, 13.40.Ks, 13.66.Jn 
\end{keyword}
\end{frontmatter}
%
\input{introduction}
\input{matching}
\input{numerics}
\input{2loop}
\input{conclusions}
\appendix
\input{appendix}
\input{bibliography}
%
\end{document}

%% file: introduction.tex
\section{Introduction}
\label{intro}
The determination of the $R$ ratio in electron-positron annihilation, 
defined as 
$R = \sigma(e^+  e^- \to {\rm hadrons})/\sigma(e^+ e^- \to \mu^+ \mu^-)$, 
is a classical measurement of particle physics and still a quantity of deep 
interest in modern research about the fundamental constituents of matter. 
The measurement of the cross section for electron-positron annihilation into
hadrons is, in fact, an important task of high-luminosity colliders
operating in the 
region of hadronic resonances, such as $\Phi$, $\tau$-charm and $B$
factories~\cite{sighad2003}.
The reason is that the $R$ value is crucial for precise predictions of
the hadronic
contribution to  $(g-2)_{\mu}$, the anomalous magnetic moment of the
muon~\cite{massimo}, and to 
the running of the electromagnetic coupling from its value at low energy up to
high energies~\cite{jeg-2003}. In particular, the QED coupling
constant evaluated at $Z$ pole 
$\alpha_{QED}(M_Z)$ is a fundamental ingredient in precision tests
of the electroweak theory and its uncertainty critically limits the
bound on the Higgs boson mass in the indirect search through fits to 
precision data~\cite{bolek}.

The $R$ value has been measured by many experiments from hadron
production threshold to the mass of the $Z$ boson. Below the bottom 
threshold and, especially, in the energy range below 5~GeV, precision
measurements of $R$ are motivated by reducing the uncertainty of the 
hadronic contribution to $(g-2)_{\mu}$ and $\alpha_{QED}(M_Z)$.
The experimental methods presently employed to determine
$R$ include direct measurements and indirect measurements of the
hadron production cross section~\cite{zhao}.  The first approach is
followed in the recent measurements by CMD-2
and SND collaborations at
VEPP-2M~\cite{cmd2snd}, BES at BEPC~\cite{bes} and CLEO at CESR~\cite{cleo}. 
The indirect measurement, which
makes use of the emission of one or more hard photons from the initial
state and is known as radiative return~\cite{radreturn}, is
presently performed by KLOE collaboration at DA$\Phi$NE~\cite{kloe}
and BABAR at PEP-II~\cite{babar}, 
and is under consideration by BELLE at KEK-B.
The radiative
return is of particular interest because it enables, in principle, to
measure $R$ over the full energy range, i.e. from
hadron production threshold to the nominal centre-of-mass energy. 

Independently from the
different types of error affecting the precision of the two methods, a common 
source of systematic uncertainty comes from the knowledge of the
collider luminosity. 
To keep under control such an uncertainty, high-precision calculations 
of the QED processes $e^+ e^-\to e^+e^-$, $\mu^+ \mu^-$, $\gamma\gamma$,
and relative Monte 
Carlo generators, are required. Among the QED processes, the
large-angle Bhabha scattering is
of particular interest because of its large cross section and
its clean experimental signature. To 
simulate the experimentally relevant distributions and calculate the
cross section of the Bhabha process, KLOE and CLEO 
collaborations make use of the QED Parton Shower generator
\BABAYAGA, developed in Refs.~\cite{babayaga,ips} with a
precision target of 0.5\%. The Monte 
Carlo $\tt MCGPJ$~\cite{mcgpj}, which includes exact $\mathcal{O}(\alpha)$ 
corrections supplemented with leading logarithmic higher-order contributions and has 
an estimated accuracy of about 0.2\%, is 
presently used at VEPP-2M to monitor the collider luminosity. To keep under control 
the theoretical precision, other codes, such as the $\mathcal{O}(\alpha)$ generator
of Ref.~\cite{bhagenf} (based on Ref.~\cite{BK}) and the
Yennie-Frautschi-Suura tool \BHWIDE~\cite{bhwide}, are also
employed by the experimental collaborations.
A generally good agreement between theory and data, as well as between the
results of the different generators, is observed, thus confirming the precision claims
of the respective calculations~\cite{kloe,mcgpj,achim}.

Nevertheless, further progress in the calculation of radiative corrections to
QED processes and, in particular, the development of precise 
large-angle Bhabha generators are still required. This is motivated by a 
number of reasons. First, the total luminosity error quoted by KLOE 
is presently 0.6\% \cite{kloe}, where the dominant source of
uncertainty comes from theory, i.e. from the 0.5\% physical precision
inherent the \BABAYAGA\ generator. The reduction of such an error
demands progress on the Bhabha theory side. Secondly, the measurement
of the hadronic cross section in the $\pi^+ \pi^-$ channel at VEPP-2M
has achieved a total systematic error of 0.6$-$1\% in the region of
the $\rho$ resonance~\cite{cmd2snd}, which requires,
in turn, an assessment of the collider luminosity at the level of
0.1\%. Last but not least, precision measurements of $R$ trough
radiative return at high-luminosity $e^+ e^-$ storage rings KEK-B and
PEP-II are already performed or foreseen in the near future, as
previously mentioned. These
facts are also among the motivations of the recent efforts in the
direction of complete two-loop calculations to Bhabha 
scattering~\cite{bfradcor2005,arbuzovsherbak,penin,boncianietal,boncianiferroglia,box}.
The need for keeping under control accurately radiative corrections to 
QED processes has been recently reinforced by the update of 
the $e^+ e^- \to \pi^+ \pi^-$ cross section by SND 
collaboration at VEPP-2M. This reanalysis \cite{snd}, which leads to a 
decrease of the measured cross section by two systematic errors in
average with respect to the previous one,
became necessary due to a flaw in
the Monte Carlo generators previously used in data analysis to 
compute radiative corrections to $e^+ e^- \to \pi^+ \pi^-$
and $e^+ e^- \to \mu^+ \mu^-$.

The aim of the present paper is to report on a high-precision
calculation of 
radiative corrections to the Bhabha process, in order to improve the theoretical
formulation of the original \BABAYAGA\ generator down to $\mathcal{O}$(0.1\%). The
approach is based on the matching of exact next-to-leading-order corrections
with resummation through all orders of $\alpha$ of the leading contributions from 
multiple soft and collinear radiation, which are taken into account according to a
QED Parton Shower.
\footnote{It is worth stressing that the matching procedure here
developed allows for a fully exclusive generation of multiple-photon
radiation. With respect to different exclusive exponentiation
approaches, such as the Yennie-Frautschi-Suura method~\cite{yfsbrighton}, the
present approach differs in some implementation details and, in
particular, in the resummation of non-infrared single collinear
logarithms. However, since both formalisms coincide at
first order, the differences start at \oatwo\ and are not infrared sensitive.}
Emphasis is also put on the impact of higher-order non-leading-log
corrections, such as $\alpha^2 L$ (with $L$ collinear
logarithm), and non-photonic light pair contributions, to arrive at a sound estimate of the 
overall theoretical error. A critical comparison of the formulation presented 
in this paper with existing two-loop calculations allows to 
put on a more quantitative ground the estimated theoretical uncertainty. 

The paper is organized as follows. 
In Sect.~\ref{matching} we describe in detail
the matching of next-to-leading-order corrections with Parton Shower,
which the old version of \BABAYAGA\ was based on. In
Sect.~\ref{pheno} the predictions for large-angle Bhabha scattering of
the improved version of \BABAYAGA\ are
compared with those of independent generators, both for integrated cross sections 
and differential distributions of experimental interest.
In Sect.~\ref{theoacc} different sources of theoretical uncertainties are 
investigated: vacuum polarization uncertainty, approximate treatment 
of two real photon emission, light pair corrections, missing 
virtual plus soft corrections to one real photon emission. 
Section~\ref{thun} is devoted to the comparison of the formulation implemented 
in \BABAYAGA, expanded at ${\ourcal O}(\alpha^2)$, 
with two existing calculations 
of next-to-next-to-leading-order corrections to Bhabha scattering: the 
purely photonic contribution (virtual plus soft-real corrections)
and the two-loop $N_f=1$ complete 
calculation in the soft-pair approximation for real pair production. 
These comparisons corroborate the claimed physical precision 
of \BABAYAGA\ of ${\ourcal O}(0.1\%)$. 
It is worth stressing that, even with a complete two-loop calculation 
at hand, the effect of higher order corrections is still relevant 
on the scale of 0.1\% accuracy, as proved in Sect.~\ref{oltre2loop}.
Conclusions and possible developments are given in Sect.~\ref{conclusion}.

%% file: matching.tex
\section{Matching next-to-leading corrections with Parton Shower} 
\label{matching}
In this Section we discuss the consistent inclusion of an exact 
fixed order calculation in a cross section resumming the leading
corrections up to all orders of perturbation theory, in a QED Parton
Shower (PS) approach. The algorithm described below is now implemented
in the new version of the event generator \BABAYAGA~\cite{sitobabayaga}, at
present only for the Bhabha process.

The matching of the two calculations is a non trivial task
and it has to avoid the double counting at first order in $\alpha$
of the leading corrections already accounted for by the PS.

A general expression for the cross section with the emission of an
arbitrary number of photons, in leading-log (LL) approximation,
can be cast in the following form:
\be
d\sigma^{\infty}_{LL}=
{\Pi}(Q^2,\varepsilon)~
\sum_{n=0}^\infty \frac{1}{n!}~|{\ourcal M}_{n,LL}|^2~d\Phi_n
\label{generalLL}
\ee
where ${\Pi}(Q^2,\varepsilon)$ is the Sudakov form-factor accounting for the
soft-photon (up to an energy equal to $\varepsilon$ in units of the
incoming fermion energy $E$) and virtual emissions, $\varepsilon$ is
an infrared separator 
dividing soft and hard radiation and $Q^2$ is
related to the energy scale of the process.
$|{\ourcal M}_{n,LL}|^2$ is the squared amplitude in LL
approximation describing the process with the emission of $n$ hard
photons, with energy larger than $\varepsilon$ in units of $E$.
$d\Phi_n$ is the exact
phase-space element of the process (divided by the incoming flux
factor), with the emission of $n$ additional photons
with respect to the Born-like final-state configuration: considering
Bhabha scattering and
defining
$p_1$, $p_2$, $p_3$, $p_4$ and $k_i$ ($i=0,\cdots, n$) as the 
initial-state $e^-$ and $e^+$, the final-state $e^-$ and $e^+$ and photons'
momenta respectively, $d\tilde{\Phi}_n\equiv d\Phi_n\times flux$ reads
\begin{equation}
d\tilde{\Phi}_n = \left\{
\begin{array}{rl}
&\frac{1}{(2\pi)^{2}}
\frac{d^3{\vec{p}_3}}{2p_3^0}\frac{d^3{\vec{p_4}}}{2p_4^0}~
\delta^4(p_1+p_2 - p_3 -p_4)~~~~~~~~~~~~~~~~~~~~~~~\mbox{if } n = 0\\
&\frac{1}{(2\pi)^{3n+2}}
\frac{d^3{\vec{p}_3}}{2p_3^0}\frac{d^3{\vec{p}_4}}{2p_4^0}~
\delta^4(p_1+p_2 - p_3 -p_4 - \sum_{i=1}^nk_i)
\prod_{i=1}^n\frac{d^3{\vec{k}_i}}{2k_i^0}
\\
&~~~~~~~~~~~~~~~~~~~~~~~~~~~~~~~~~~~~~~~~~~~~~~~~~~~~~~~~~~
~~~~~~~~~\mbox{if } n>0
\end{array}
\right.
\end{equation}
The cross section $d\sigma^{\infty}_{LL}$ of Eq.~\myref{generalLL}
is independent of the infrared separator $\varepsilon$, provided it
is sufficiently small.

According to the factorization theorems of soft and/or collinear
singularities,
the squared amplitudes in LL approximation can be written in a factorized
form. In the following, for the sake of clarity and without loss of
generality, we write photon emission formulas as if only one external
fermion radiates. We are aware that it is a completely unphysical
case, but it allows to write more compact formulas, being the
generalization to the real case straightforward when including the
suited combinatorial factors. With this in mind,
the one-photon emission squared amplitude in LL approximation can be
written as
\be
|{\ourcal M}_{1,LL}|^2=\frac{\alpha}{2\pi}\frac{1+z^2}{1-z}I(k)
~|{\ourcal M}_0|^2~
\frac{8\pi^2}{E^2 z (1-z)}
\label{onegammaLL}
\ee
where $1-z$ is the fraction of the fermion energy $E$ carried by the
photon, $k$ is the photon four-momentum,
$I(k)$ is a function describing the angular spectrum of the photon
and $P(z)=(1+z^2)/(1-z)$ is the Altarelli-Parisi 
${e} \to {e} + \gamma$ splitting function. 
In Eq.~\myref{onegammaLL} we observe the factorization of the Born squared
amplitude and that the emission factor $\frac{\alpha}{2\pi}P(z)I(k)
\frac{8\pi^2}{E^2z(1-z)}$
can be iterated for each photon emission, up to all orders,
to obtain $|{\ourcal M}_{n,LL}|^2$.
It is worth noticing that ${d^3\vec{k}}/{k^0} = (1-z)E^2d\Omega_\gamma dz$
and that in the collinear limit the cross
section of Eq.~\myref{generalLL} reduces to the cross section calculated
by means of the QED PS algorithm described in Refs.~\cite{babayaga,ips}.

The Sudakov form factor $\Pi(Q^2,\varepsilon)$ reads explicitly
\be
\Pi(Q^2,\varepsilon)=
\exp\left(
-\frac{\alpha}{2\pi}~I_+~L^\prime
\right),~~~~L^\prime=\log\frac{Q^2}{m^2},~~~~
I_+\equiv
\int_0^{1-\varepsilon}
dz P(z)
\ee
The function $I(k)$ has the property that
$
\int d\Omega_{\gamma}I(k)=\log Q^2/m^2
$
and allows the cancellation of the infrared logarithms.

The cross section calculated in Eq.~\myref{generalLL} has the
advantage that the photonic corrections, in LL approximation, are
resummed up to all orders of perturbation theory. On the other side,
the weak point of the formula~\myref{generalLL} is that its
expansion at \oa\ does not coincide with an exact \oa\ (NLO) result, being
its LL approximation. In fact
\bea
d\sigma^{\alpha}_{LL}&=&
\left[
1-\frac{\alpha}{2\pi}~I_+~\log\frac{Q^2}{m^2}
\right] |{\ourcal M}_0|^2 d\Phi_0+
|{\ourcal M}_{1,LL} |^2 d\Phi_1\nonumber\\
&\equiv&
\left[
1+C_{\alpha,LL}
\right] |{\ourcal M}_0|^2 d\Phi_0
+
|{\ourcal M}_{1,LL} |^2 d\Phi_1
\label{LL1}
\eea
whereas an exact NLO cross section can be always cast in the form
\be
d\sigma^{\alpha}
=
\left[
1+C_{\alpha}
\right] |{\ourcal M}_0|^2 d\Phi_0
+
|{\ourcal M}_{1} |^2 d\Phi_1
\label{exact1}
\ee
The coefficient $C_{\alpha}$ contains the complete virtual
\oa\ and the \oa\ soft-bremsstrahlung squared matrix elements, in units of
the Born squared amplitude,
and $|{\ourcal M}_1|^2$ is the exact squared matrix element with the
emission of one hard photon.
We remark that $C_{\alpha,LL}$ has the same logarithmic structure as
$C_\alpha$ and that $|{\ourcal M}_{1,LL}|^2$ has the same singular
behaviour of $|{\ourcal M}_1|^2$.

In order to match the LL and NLO calculations, we introduce
the correction factors, which are by construction
infrared safe and free of collinear logarithms,
\be
F_{SV}~=~
1+\left(C_\alpha-C_{\alpha,LL}\right),~~~~~~
F_H~=~
1+\frac{|{\ourcal M}_1|^2-|{\ourcal M}_{1,LL}|^2}{|{\ourcal M}_{1,LL}|^2}
\label{FSVH}
\ee
and we notice that the exact \oa\ cross section can be expressed, up
to terms of ${\ourcal O}(\alpha^2)$,
in terms of its LL approximation as
\be
d\sigma^\alpha~=~
F_{SV} (1+C_{\alpha,LL} ) |{\ourcal M}_0|^2 d\Phi_0
~+~
F_H |{\ourcal M}_{1,LL}|^2 d\Phi_1
\label{matchedalpha}
\ee
Driven by Eq.~\myref{matchedalpha}, Eq.~\myref{generalLL} can be improved
by writing the resummed cross section as
\be
d\sigma^{\infty}_{matched}=
F_{SV}~\Pi(Q^2,\varepsilon)~
\sum_{n=0}^\infty \frac{1}{n!}~
\left( \prod_{i=0}^n F_{H,i}\right)~
|{\ourcal M}_{n,LL}|^2~
d\Phi_n
\label{matchedinfty}
\ee
The correction factors $F_{H,i}$ follow from the definition Eq.~\myref{FSVH}
for each photon emission.
The expansion at \oa\ of Eq.~\myref{matchedinfty} coincides now with
the exact NLO cross section Eq.~\myref{exact1} and
all higher order LL contributions
are the same as in Eq.~\myref{generalLL}.
%

Eq.~\myref{matchedinfty} is our master formula for the matching between
the exact \oa\ calculation and the QED resummed PS
cross section, according to which we also generate events.
The extension of the matching formula Eq.~\myref{matchedinfty} to the
realistic case, where every charged particle radiates photons,
is almost straightforward. We would like to remark also that
the LL cross section of Eq.~\myref{generalLL} is by construction
positively defined in every point of the phase space, whereas the
correction factors of Eq.~\myref{FSVH} can in principle make the
differential cross section of Eq.~\myref{matchedinfty} negative in
some point, namely where the PS approximation is less accurate
(e.g. for hard photons at large angles). Nevertheless, we verified
that this never happens when considering typical event selection criteria for
luminosity at flavour factories.

It is useful to present, in the realistic case, the expression of the
function $I(k)$, which describes the leading behaviour of the angular spectrum
of the emitted photons, accounting also for interference of radiation
coming from different charged particles:
\be
I(k)~=~
\sum_{i,j=1}^4~
\eta_i \eta_j~
\frac{p_i\cdot p_j}{(p_i\cdot k)(p_j\cdot k)}~
E_\gamma^2
\label{idik}
\ee
where $p_l$ is the momentum of the external fermion $l$, 
$\eta_l$ is a charge factor equal to +1
for incoming particles or outgoing antiparticles
and equal to -1 for incoming antiparticles or outgoing particles,
$k$ is the photon momentum, $E_\gamma$ is its energy
and the sum runs over all the external fermions. The function $I(k)$
does not depend on the photon energy. Given Eq.~\myref{idik}, one can
quite easily convince that the more convenient choice for the Sudakov form
factor scale $Q^2$ is $L^\prime=\log\frac{Q^2}{m^2} =
\log\frac{st}{u m^2} - 1\equiv L-1$ where $s$, $t$ and $u$ are the
Mandelstam variables of the process and $m$ is the electron mass.

The exact squared amplitude for the emission of a real photon $|{\ourcal M}_1|^2$ 
has been calculated by hand with the help of $\tt FORM$~\cite{vermaseren} and
successfully cross checked with Ref.~\cite{BK} and with the output of
the $\tt ALPHA$ algorithm~\cite{alpha}.

The exact \oa\ soft plus virtual corrections to the Bhabha scattering
have been taken from Ref.~\cite{cafforemiddi}. The soft plus virtual
cross section reads
\bea
d\sigma^\alpha_{SV} &=&
d\sigma^{\alpha,s}_{SV}+d\sigma^{\alpha,t}_{SV}
+d\sigma^{\alpha,st}_{SV}\nonumber\\
d\sigma^{\alpha,i}_{SV}&=&d\sigma_0^i [2 (\beta + \beta_{int})
\log\varepsilon+C_F^i]
\label{exactsv}
\eea
where $i$ is an index for $s$, $t$ and $s$-$t$ subprocesses 
contributing to the 
Bhabha cross section, $\beta = \frac{2\alpha}\pi[\log(s/m^2)-1]$, $\beta_{int} =
\frac{2\alpha}\pi\log(t/u)$ and the explicit expression for $C_F^i$
can be found in Ref.~\cite{cafforemiddi}. We notice that in
Eq.~\myref{exactsv} the terms coming from $s$, $t$ and $s$-$t$
interference diagrams are explicitly given.

In the following, we discuss 
the implementation of the vacuum polarization effects in \BABAYAGA.
Some 
technical aspects about
Eq.~\myref{matchedinfty} (its independence from $\varepsilon$,
the mapping of the momenta needed for $n\ge 2$ and the importance sampling of
the final-state collinear singularities) are discussed in
Appendix~\ref{appendice}.

\subsection{Vacuum polarization and $Z$ exchange contributions}
\label{vpsubsection}
Besides the photonic radiative corrections 
considered in the previous Section, also the vacuum
polarization effects must be included in the master 
formula~\myref{matchedinfty}, in order to reach the required theoretical accuracy. 
They are accounted for by replacing
the fine structure constant $\alpha\equiv\alpha(0)$ with 
$\alpha(q^2)=\alpha/(1-\Delta\alpha(q^2))$, where $\Delta\alpha(q^2)$
is the fermionic contribution to the photon self-energy: the leptonic
and top-quark one-loop contributions can be calculated analytically in
perturbation theory, while the remaining five quarks (hadronic)
contribution, $\Delta\alpha^{(5)}_{hadr}$, has to be extracted from
data. To evaluate $\Delta\alpha^{(5)}_{hadr}$ we use the $\tt HADR5N$
routine by F. Jegerlehner~\cite{jeg-2003,hadr5}.

Setting $r_s=\alpha(s)/\alpha$ and $r_t=\alpha(t)/\alpha$ ($s$ and $t$
are the Mandelstam invariants), we include vacuum polarization
in the Born matrix element, which is proportional to $\alpha$, by
rescaling the $s$ and $t$ channel amplitudes, namely
\be
|{\ourcal M}_0|^2=|{\ourcal M}_{0,s} + {\ourcal M}_{0,t}|^2~~\to~~
|{\ourcal M}_{0,VP}|^2 = |{\ourcal M}_{0,s}r_s + {\ourcal M}_{0,t}r_t|^2
\label{vpborn}
\ee

Going beyond the Born-like approximation, the cross section corrected
at \oa\ including also vacuum polarization can be written as
$\sigma^\alpha_{VP}=\sigma_{0,VP} + \sigma^\alpha_{SV} +
\sigma^\alpha_{H}$, where $\sigma^\alpha_{SV}$ and $\sigma^\alpha_{H}$
are the soft plus virtual and the hard photon \oa\ corrections of
photonic origin. We can go a step further and include vacuum
polarization in those terms, in order to include also part of the
${\ourcal O}(\alpha^2)$ factorizable corrections. The hard emission
matrix element is the sum of eight amplitudes where the real photon is
attached to a $s$ or $t$ channel-like diagram. As in
Eq.~\myref{vpborn}, we rescale those amplitudes by $r_s$ and $r_t$,
respectively. In order to guarantee the cancellation of the infrared
separator $\varepsilon$ between $\sigma^\alpha_{SV}$ and
$\sigma^\alpha_{H}$, also in $\sigma^\alpha_{SV}$ of Eq.~\myref{exactsv}, 
we rescale the $s$,
$t$ and $s$-$t$ interference contributions with the appropriate vacuum 
polarization factor, namely $r_s^2$, $r_t^2$ and $r_sr_t$.
Finally, the vacuum polarization improved amplitudes and cross
sections are used as building blocks of the master 
formula~\myref{matchedinfty}.

Furthermore, we add to the Born amplitude also the $Z$ exchange diagrams: their
effect is really tiny and negligible at low energies, but can become
more important (up to 0.1\%) around 10 GeV when considering wide
angular acceptance regions.

%

%% file: numerics.tex
\section{Numerical results}
\label{pheno}
In order to test the internal consistency of the formulation described above 
and to quantify the physical precision of the improved version of the 
\BABAYAGA\ generator, 
we performed a number of Monte Carlo simulations of those experimental observables 
which are relevant for luminosity measurements at $e^+ e^-$ flavour
factories. To model the acceptance
cuts used by the experimental collaborations, we considered four different set up 
defined by the following selection criteria
\begin{enumerate}
\renewcommand{\theenumi}{\alph{enumi}}
\item $\sqrt{s}=1.02~{\rm GeV},\; E_{min} = 0.408~{\rm GeV},\;
20^{\circ} < \theta_\pm < 160^{\circ},\; \xi_{max}=10^{\circ}$

\item $\sqrt{s}=1.02~{\rm GeV},\; E_{min}=0.408~{\rm GeV},\;
55^{\circ} < \theta_\pm < 125^{\circ},\; \xi_{max}=10^{\circ}$

\item $\sqrt{s}=10~{\rm GeV},\; E_{min}=4~{\rm GeV},\; 
20^{\circ} < \theta_\pm < 160^{\circ},\; \xi_{max}=10^{\circ}$

\item $\sqrt{s}=10~{\rm GeV},\; E_{min}=4~{\rm GeV},\; 
55^{\circ} < \theta_\pm < 125^{\circ},\; \xi_{max}=10^{\circ}$
\end{enumerate}
where $E_{min} = 0.8\times E_{beam}$ is the energy threshold for the 
final-state electron/positron, $\theta_{\pm}$ are the electron/positron scattering
angles and $\xi_{max}$ is the maximum allowed acollinearity. The set up (a) and
(b) are of interest for $\Phi$ factories, while set up (c) and (d) refer to $B$-factories. In
both cases, a wider and a tighter angular acceptance are considered, in order to study
the dependence of the radiative corrections from the acceptance criteria. The energy and
acollinearity cuts are very similar to those considered in previous simulations and tend to
single out quasi-elastic Bhabha events. It is understood that the energy of final-state electron/positron
corresponds to a so-called ``bare'' event selection
(i.e. without photon recombination), which resembles
realistic data taking at flavour factories.

\subsection{Integrated cross sections and technical tests}
\begin{figure}[ht]
 \begin{center}
\includegraphics{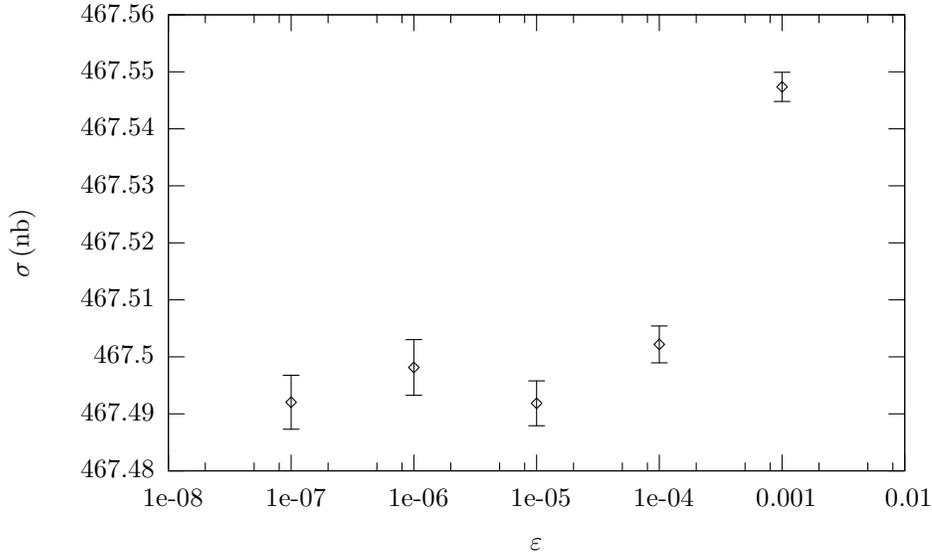}
 \caption{QED corrected Bhabha cross section as a function of the infrared regulator 
 $\varepsilon$, according to set up (b). The error bars correspond to
 $1\sigma$ Monte Carlo error.}
 \label{fig:fig1}
 \end{center}
 \end{figure}

A first meaningful test of the correct matching of NLO corrections with PS is to
prove independence of the predictions for the QED corrected cross section from variation
of the soft-hard separator $\varepsilon$ required by the PS algorithm. This is successfully demonstrated
in Fig.~\ref{fig:fig1}, which shows the Bhabha cross section, obtained 
according to Eq.~\myref{matchedinfty} and in the 
conditions of set up (b), as a function of $\varepsilon$ from $10^{-3}$ to $10^{-7}$. A priori, 
one should expect compatibility of the calculated cross section against 
$\varepsilon$ variation at an accuracy level of $\mathcal{O}(\alpha \varepsilon)$. 
This is clearly seen to be valid in Fig.~\ref{fig:fig1}, 
when looking at the relative difference between the cross section predictions as
$\varepsilon$ varies. Also for the $\mathcal{O}(\alpha)$ corrected cross section, which is
an important ingredient of the present formulation and a component of the following 
discussion, independence from $\varepsilon$  has been successfully checked in the limit of
sufficiently small values, i.e. for $\varepsilon$ variations in the range $10^{-3}-10^{-7}$. 

To quantify the overall impact of QED radiation and, in particular, to evaluate the size of
QED contributions at different perturbative/precision levels, we show in Tab.~\ref{table:sigmas}
the lowest-order Bhabha cross section (without and with vacuum polarization), the 
exact $\mathcal{O}(\alpha)$ cross section as in Eq.~\myref{exact1} and the 
$\mathcal{O}(\alpha)$ PS cross section as 
in Eq.~\myref{LL1}, as well as the all-order PS cross section of Eq.~\myref{generalLL} 
and the matched PS cross section 
of Eq.~\myref{matchedinfty}. 
The four columns correspond to the experimental conditions previously
defined. In the cross sections of Tab.~\ref{table:sigmas}, we switch
off the vacuum polarization effect except in the second row, to better
study the different sources of corrections. Interestingly, 
from these cross section values it is possible to disentangle the relative effect of various 
QED contributions, namely the contribution of vacuum polarization, of exact 
$\mathcal{O}(\alpha)$ corrections, of higher-order (i.e. beyond $\mathcal{O}(\alpha)$)
leading corrections in the pure PS approach and in the improved PS matched with
$\mathcal{O}(\alpha)$ corrections, of non-logarithmic terms entering the
$\mathcal{O}(\alpha)$ cross section and present also in the improved PS algorithm and, 
finally, of part of the sub-leading $\alpha^n L^{n-1}$ effects. 
The above per cent corrections are shown in Tab. \ref{table:deltas}
and they can be derived from the cross section values of
Tab.~\ref{table:sigmas} according to the following formulae 
\begin{eqnarray*}
   \delta_{VP}       &\equiv& \frac{\sigma_{0,VP}-\sigma_{0}}{\sigma_{0}}~~~~~~~~~~~~~~~~~~~~~~~~~~~
   \delta_{\alpha}    \equiv  \frac{\sigma^{NLO}_\alpha-\sigma_0}{\sigma_0}\nonumber\\
   \delta_{HO}       &\equiv& \frac{\sigma^{PS}_{matched}-\sigma_{\alpha}^{NLO}}{\sigma_{0}}~~~~~~~~~~~~~~~~~~
   \delta_{HO}^{PS}   \equiv  \frac{\sigma^{PS} - \sigma^{PS}_\alpha}{\sigma_{0}}\nonumber\\
   \delta_{\alpha}^{non\mbox{-}log} &\equiv&
                              \frac{\sigma^{NLO}_\alpha-\sigma^{PS}_\alpha}{\sigma_{0}}~~~~~~~~~~~~~~~~~~~
   \delta_{\infty}^{non\mbox{-}log}  \equiv 
                              \frac{\sigma^{PS}_{matched}-\sigma^{PS}}{\sigma_{0}}\nonumber\\
   \delta_{\alpha^2L}&\equiv& \frac{\sigma^{PS}_{matched}-\sigma^{NLO}_\alpha -\sigma^{PS}+\sigma^{PS}_\alpha}
                                   {\sigma_{0}} \nonumber
\end{eqnarray*}
\begin{table}[thb]
\begin{center}
\begin{tabular}{|p{2.775cm}|c|c|c|c|}
      \hline
      set up               &     (a)   &
      (b)   &      (c)   &     (d) \\
      \hline
      \hline
      $\sigma_{0}   $      &
      $6855.743(1)$      &     $529.4631(2)$  &
      $71.333(1)$   &     $5.5026(2)$    \\
      $\sigma_{0,VP}$      &
      $6976.49(4)$   &     $542.657(6)$  &
      $74.7632(6)$   &     $5.85526(3)$    \\
      $\sigma^{NLO}_\alpha$     &
      $6060.07\left (6\right )$ &     $451.523\left (6\right )$  &
      $59.900\left (1\right )$  &     $4.4256\left (2\right )$    \\
      $\sigma^{PS}_\alpha$      &
      $6083.59\left (6\right )$ &     $454.503\left (6\right )$  &
      $60.144\left (1\right )$  &     $4.4565\left (1\right )$    \\
      $\sigma_{matched}^{PS}$      &
      $6086.74\left (7\right )$   &     $455.858\left (5\right )$  &
      $60.419\left (1\right )$   &     $4.5046\left (3\right )$    \\
      $\sigma^{PS}$      &
      $6107.57\left (6\right )$   &     $458.437\left (4\right )$  &
      $60.628\left (1\right )$   &     $4.5301\left (2\right )$    \\
      \hline
\end{tabular}
\end{center}
\caption{Bhabha cross section (in nb) according to different precision levels and for
the four set up specified in the text.}
\label{table:sigmas}
\end{table}
\begin{table}[thb]
\begin{center}
\begin{tabular}{|p{2.775cm}|c|c|c|c|}
      \hline
      set up &(a)&(b)&(c)&(d)\\
      \hline      
      \hline
      $\delta_{VP}$                     &
      $1.76$         &     $2.49$       &
      $4.81$         &     $6.41$       \\
      $\delta_{\alpha}$                 &
      $-11.61$       &     $-14.72$     &
      $-16.03$       &     $-19.57$     \\
      $\delta_{HO}$                     &
      $0.39$        &     $0.82$        &
      $0.73$        &     $1.44$        \\
      $\delta_{HO}^{PS}$                &
      $0.35$        &     $0.74$        &
      $0.68$        &     $1.34$        \\
      $\delta_{\alpha^{2}L}$            &
      $0.04$        &     $0.08$        &
      $0.05$        &     $0.10$        \\
      $\delta_{\alpha}^{non\mbox{-}log}$&
      $-0.34$       &     $-0.56$       &
      $-0.34$       &     $-0.56$       \\
      $\delta_{\infty}^{non\mbox{-}log}$&
      $-0.30$       &     $-0.49$       &
      $-0.29$       &     $-0.46$       \\
      \hline
\end{tabular}
\end{center}
\caption{Relative corrections (in per cent) to the Bhabha cross
section for the four set up specified in the text.}
\label{table:deltas}
\end{table}

From Tab. \ref{table:deltas} it can be seen that the vacuum polarization gives a positive 
correction to the lowest-order cross section of the order of 2\% at the $\Phi$ factories 
and of about 5-6\% at the $B$ factories, being its contribution more important for a tighter
angular acceptance than for a wider one. This dependence of vacuum polarization 
from the detector acceptance has to be ascribed to the role played by the different 
sub-processes contributing to the Bhabha cross section as the angular acceptance varies. 
Actually, the energy scale entering the logarithmic dependence of the vacuum polarization
is process dependent and is equal to the c.m. energy $s$ for the time-like $s$-channel
sub-process and to (the absolute value of) the momentum transfer $|t|$ for the 
space-like $t$-channel contribution, being on the average $|t| \ll s$. While for a wide acceptance the $t$-channel contribution
is largely dominating, the $s$-channel contribution becomes more and more important as
the angular acceptance decreases, thus explaining the trend observed for the vacuum 
polarization correction. The exact $\mathcal{O}(\alpha)$ corrections
lower the cross section of about 15\% ($\Phi$ factories) and of about 20-25\% ($B$-factories).
Higher-order contributions of the type $\mathcal{O}(\alpha^n L^n)$, with $n \ge 2$, introduce
a positive correction around 0.5-1\% at the $\Phi$ factories and at the 1-2\% level at the
$B$ factories. Therefore, multiple photon corrections are unavoidable in view of the 
required theoretical precision, as already noticed in Ref.~\cite{babayaga}. 
On the other hand, also non-log 
$\mathcal{O}(\alpha)$ corrections are necessary at a precision level of 0.1\%, since their
contribution is of the order of 0.5\%, almost independently from the c.m. energy and with
a mild dependence from the angular cuts. This confirms {\em a
  posteriori} the need for matching
the original PS formulation of \BABAYAGA\ with NLO corrections. The
effect due to
$\mathcal{O}(\alpha^2 L)$ corrections varies from 0.05\% (wide acceptance) to 
0.1\% (tight acceptance). Although these contributions are only approximately kept under
control, it can be argued, and will be shown in the following Section, that the infrared part 
of $\mathcal{O}(\alpha^2 L)$ terms is correctly reproduced
by the present approach~\cite{a2l} and, therefore, the size of such effects can be viewed as an estimate of the 
overall physical precision, which is conservatively close to 0.1\%. From Tab.~\ref{table:deltas} 
it can be also seen that the matching of NLO corrections with PS does not alter at the level of 0.1\%
the size of higher-order and NLO contributions, thus preserving correctly the impact of
these partial effects. This conclusion can be inferred by comparing $\delta_{HO}$ with 
$\delta_{HO}^{PS}$ and $\delta_{\alpha}^{non\mbox{-}log}$ with $\delta_{\infty}^{non\mbox{-}log}$, respectively.
A common feature observed for all the relative corrections, with the exception  of 
non-log effects, is the fact that, at a fixed c.m. energy, they are larger (by about a factor of two) 
for a tighter acceptance with respect to a wider one. This can be understood as follows.
As discussed above, a natural choice for the $Q^2$ dependence of the collinear logarithm 
$L = \log Q^2/m^2$ is $Q^2 = s t /u$. This implies that at large
scattering angles the collinear logarithm
tends to $\log s/m^2$, while for a wide angular acceptance 
it can be on the average approximated by
$\log |t|/m^2$, thus explaining the observed angular-dependent behaviour.

\subsection{Differential distributions}
\begin{figure}[ht]
 \begin{center}
\includegraphics{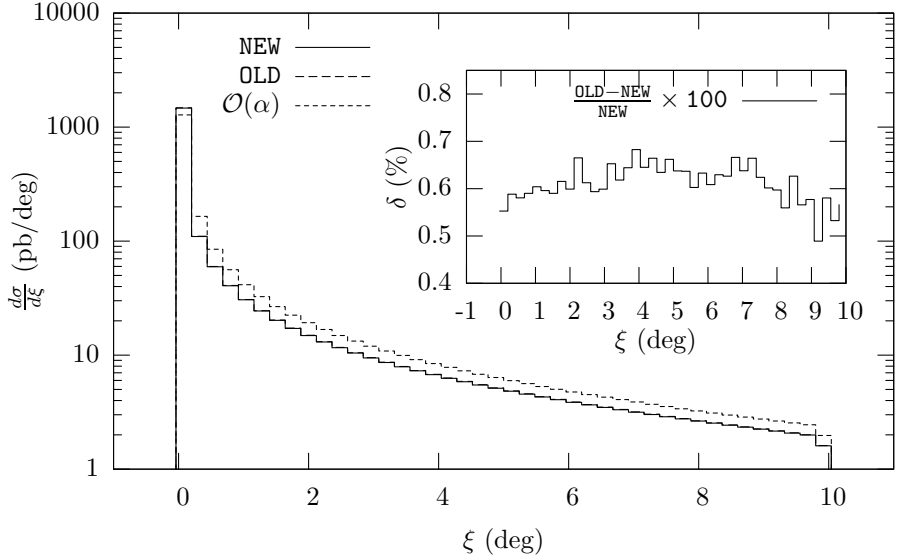}
 \caption{Acollinearity distribution according to the PS matched with $\mathcal{O}(\alpha)$ 
 corrections (Eq.~\myref{matchedinfty}, solid line), the LL PS algorithm 
 (Eq.~\myref{generalLL}, dashed line) and the exact $\mathcal{O}(\alpha)$ calculation 
 (dotted line). The inset shows the relative differences between the predictions of the 
 improved and the LL PS. Selection criteria of set up (b) are considered.}
 \label{fig:fig2}
 \end{center}
 \end{figure}

\begin{figure}[ht]
 \begin{center}
\includegraphics{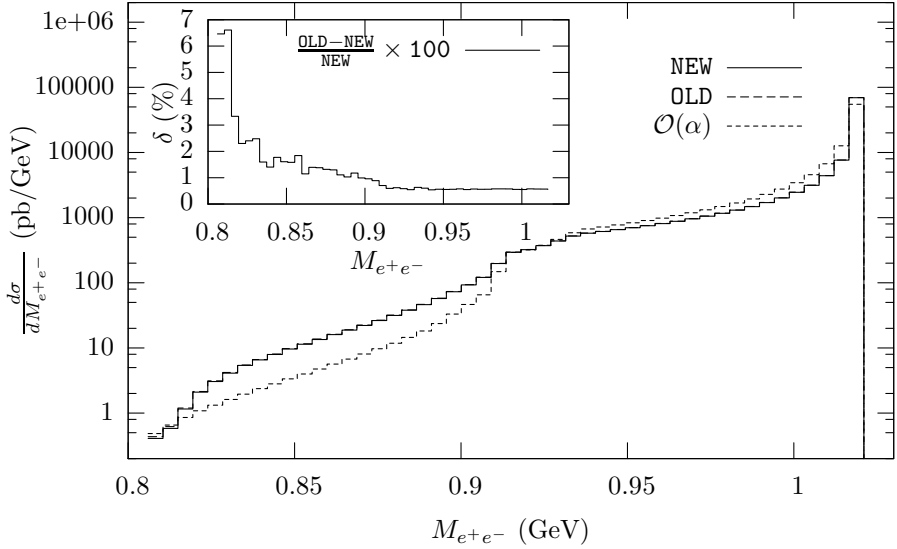}
 \caption{The same as Fig.~\ref{fig:fig2} for the $e^+ e^-$ invariant mass distribution.}
 \label{fig:fig3}
 \end{center}
 \end{figure}

Besides the integrated cross section, various differential distributions are used by the
experimental collaborations to monitor the collider luminosity. We
show, in Fig.~\ref{fig:fig2} and Fig.~\ref{fig:fig3}, two
distributions which are particularly sensitive to the details of
photon radiation, i.e. the $e^+ e^-$ acollinearity and the invariant mass
distribution, in order to quantify the differences between the
previous version of \BABAYAGA\ (denoted as $\tt  OLD$ in the figures)
and the improved one presented here (denoted as $\tt NEW$). As a reference, the distributions
obtained according to the exact $\mathcal{O}(\alpha)$ calculation are also shown.
The results refer to set up (b). 

From Fig.~\ref{fig:fig2} and Fig.~\ref{fig:fig3} it can be clearly
seen that multiple photon
corrections introduce significant deviations with respect to 
an $\mathcal{O}(\alpha)$ simulation, especially in the hard tails of
the distributions, where they amount to 
several per cent. To make more visible the improvements 
introduced by the matching procedure discussed in
Sect.~\ref{matching}, the inset shows the relative 
differences between the predictions of the improved and original
version of \BABAYAGA. 
These differences mainly come from non-log $\mathcal{O}(\alpha)$  contributions and, 
to a smaller extent, from $\mathcal{O}(\alpha^2 L)$ terms. Their effect is flat and at 
level of 0.5\% for the acollinearity distribution, while they reach the some per cent level 
in the hard tail of the invariant mass distribution. As a whole, these results demonstrate 
that exact $\mathcal{O}(\alpha)$ and higher-order corrections need to be simultaneously
taken into account for precision luminosity studies.

\subsection{Tuned comparisons}
\label{tunedcomparisons}
An important step towards the estimate of the theoretical accuracy of the formulation 
is the tuned comparison of the improved version of \BABAYAGA\ with independent 
precision calculations of the Bhabha process. To this end, we first compared the results for 
the integrated cross section as obtained by the improved version of
\BABAYAGA\ with the
corresponding predictions of \LABSPV~\cite{labspv} and
\BHWIDE~\cite{bhwide}, which both rely on different theoretical
ingredients. For the sake of comparison, the contribution of vacuum polarization has 
been switched off, in order to test just the implementation of pure QED corrections. The
comparison is shown in Tab.~\ref{table:comparison}, for both set up (a) and (b). As can be 
seen, the predictions of the three calculations agree within 0.1\%. Although we didn't 
performed detailed tests in comparison with the recently developed
generator $\tt MCGPJ$~\cite{mcgpj},
we expect a level of agreement with that calculation at a similar precision level, on the
basis of the comparisons between \BHWIDE\ and $\tt MCGPJ$ reported in Ref.~\cite{mcgpj}. 

\begin{table}[thb]
\begin{center}
\begin{tabular}{|l|c|c|}
      \hline
set up &     (a)    &      (b)   \\
      \hline
      \hline
      $\sigma_{\tt BHWIDE}$                                &
      $6086.3\left (2\right )$       &     $455.73\left (1\right )$    \\
      $\sigma_{\tt LABSPV}$                          &
      $6088.5\left (3\right )$       &     $456.19\left (1\right )$         \\
      $\sigma_{{\tt BABAYAGA}}^{matched}$      &
      $6086.61\left (2\right )$       &     $455.853\left (4\right )$  \\
      \hline
\end{tabular}
\end{center}
\caption{Comparison between the predictions for the Bhabha cross section (in nb)
as obtained with \BHWIDE, \LABSPV\ and the present version of \BABAYAGA.}
\label{table:comparison}
\end{table}

\begin{figure}[ht]
\begin{center}
\includegraphics{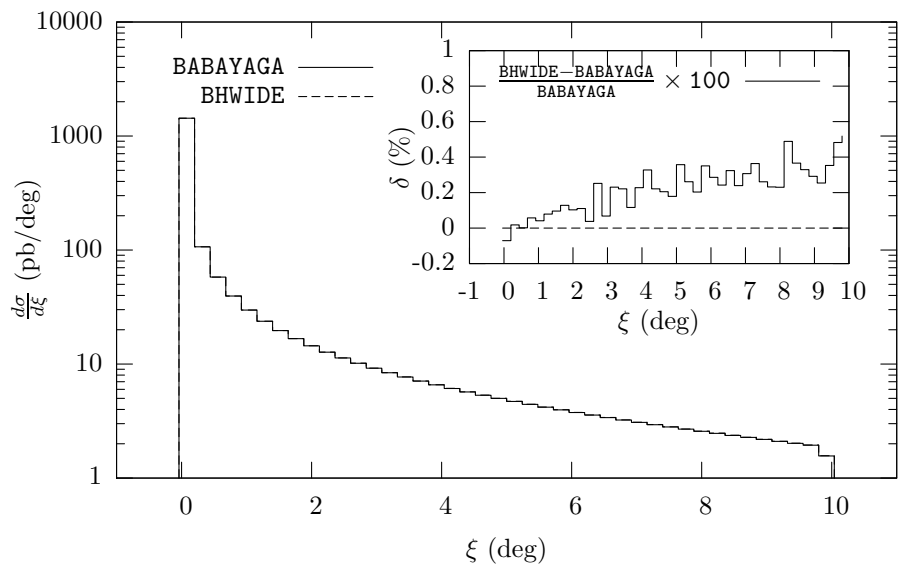}
 \caption{Comparison between the \BHWIDE\ generator (dashed line) and the present version of
 \BABAYAGA\ (solid line) for the acollinearity distribution. The inset
 shows the relative difference 
 between the predictions of the two generators. Selection criteria of set up (b) are considered.}
 \label{fig:fig4}
 \end{center}
 \end{figure}

\begin{figure}[ht]
 \begin{center}
\includegraphics{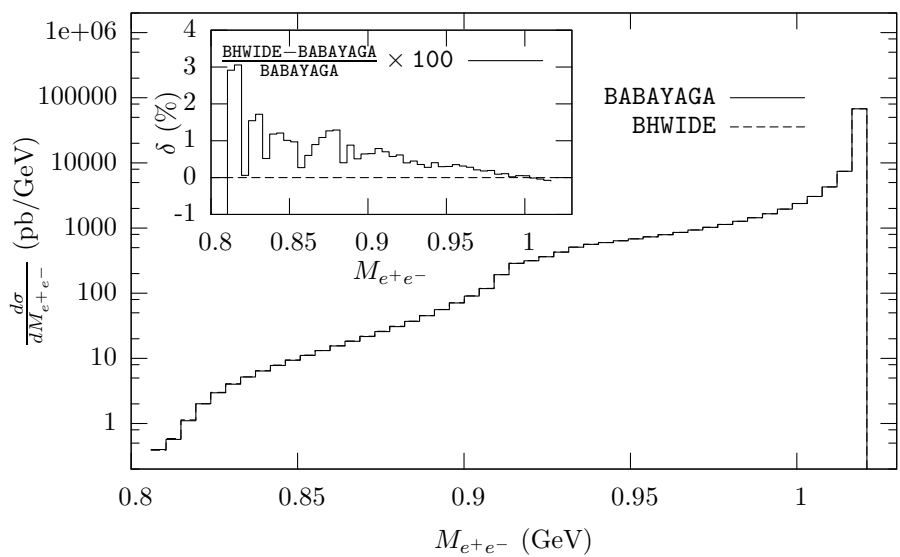}
 \caption{The same as Fig.~\ref{fig:fig4} for the $e^+ e^-$ invariant mass distribution.}
 \label{fig:fig5}
 \end{center}
 \end{figure}

Comparisons between \BABAYAGA\ and \BHWIDE\ at the level of differential distributions are
given in Fig.~\ref{fig:fig4} and Fig.~\ref{fig:fig5}, where the inset shows the relative deviations
between the predictions of the two codes, with reference to set up~(b). As can be seen, there is
a very good agreement between the two generators, as the predicted distributions 
appear, at a first sight, almost indistinguishable. Looking in more detail, there is a 
relative difference of a few per mille for the acollinearity
distribution (Fig.~\ref{fig:fig4}) and
of a few per cent for the invariant mass  (Fig.~\ref{fig:fig5}), but
only in the hard tails, 
which little contribute to the integrated cross section. In fact, these differences on differential 
distributions translate into an agreement on the cross section values well below the 0.1\% level, as
shown in Tab.~\ref{table:comparison}.  

On the ground of these results, we can conclude that the calculation
of QED corrections in the two generators, although based on different
approaches, numerically agrees very well, at the level of the required
precision for accurate luminosity measurements at flavour factories.

%% file: 2loop.tex
\section{Estimate of the theoretical accuracy}
\label{theoacc}
Since different implementations of radiative corrections beyond exact 
${\ourcal O}(\alpha)$ contributions differ by higher order effects, 
the results of the comparisons quoted in the previous section between 
\BABAYAGA\ and other event generators give 
a hint of the missing radiative corrections which dominate the
theoretical accuracy. From the investigations of the previous Section, 
related to the higher order terms inherent in the theoretical
formulation of \BABAYAGA\ and to tuned comparisons with independent
codes, an estimate of the physical precision in the calculation of
the radiative corrections of the order of 0.1\% can be inferred.

This guess can be verified by a comparison with a complete
two-loop calculation, since all neglected
terms in the formulation described in Sect.~\ref{matching} start at
two-loop order. Being a full two-loop calculation not presently available, 
the aim of the present Section is to try to estimate the impact of the
uncertainties at order $\alpha^2$ within the realistic set up
considered in this paper, by comparing with some of the available
calculations in the literature.

Another important source of error is the
uncertainty on the hadronic contribution to the vacuum polarization
$\Delta\alpha^{(5)}_{hadr}$, which will be quantified in
Sect.~\ref{vperror}. The origin of this error is intrinsically
non-perturbative because $\Delta\alpha^{(5)}_{hadr}$ can not be
calculated perturbatively around the hadronic resonances and must be
calculated via dispersion relations by means of
data~\cite{jeg-2003}. As discussed in Sec.~\ref{vperror}, this error
will be the dominant one close to the $J/\Psi$ resonances.

We start by considering the theoretical error of perturbative origin,
at \oatwo.
The pure \oatwo\ content of our master formula \myref{matchedinfty}
can be read by expanding it and can be cast in the following form
\be
\sigma^{\alpha^2} = \sigma^{\alpha^2}_{SV} + \sigma^{\alpha^2}_{SV,H} +
\sigma^{\alpha^2}_{H,H} 
\label{alpha2content}
\ee
where $\sigma^{\alpha^2}_{SV}$ contains all the
$\alpha^2$ virtual and real contributions without photons with energy
larger than $\varepsilon$, $\sigma^{\alpha^2}_{SV,H}$ contains all the
virtual and real contributions with at least one photon with energy
larger than $\varepsilon$ and $\sigma^{\alpha^2}_{H,H}$ is the
contribution with two real photons with energy larger than
$\varepsilon$. In \BABAYAGA, each of the three terms is affected in
principle by an error and in the following we try to study the impact
of this error on the integrated cross section to establish the
theoretical accuracy of the master formula~\myref{matchedinfty}.

\subsection{Corrections beyond two-loop}
\label{oltre2loop}
Before starting to discuss the
theoretical error, we would like to prove that the LL radiative corrections
beyond $\alpha^2$ are still very important, at least when considering
differential distributions. This means that even if the complete two-loop
perturbative calculation will be fully available, it would be
desirable a matching with the resummation of all the remaining
LL corrections. 

\begin{figure}[th]
\begin{center}
\includegraphics{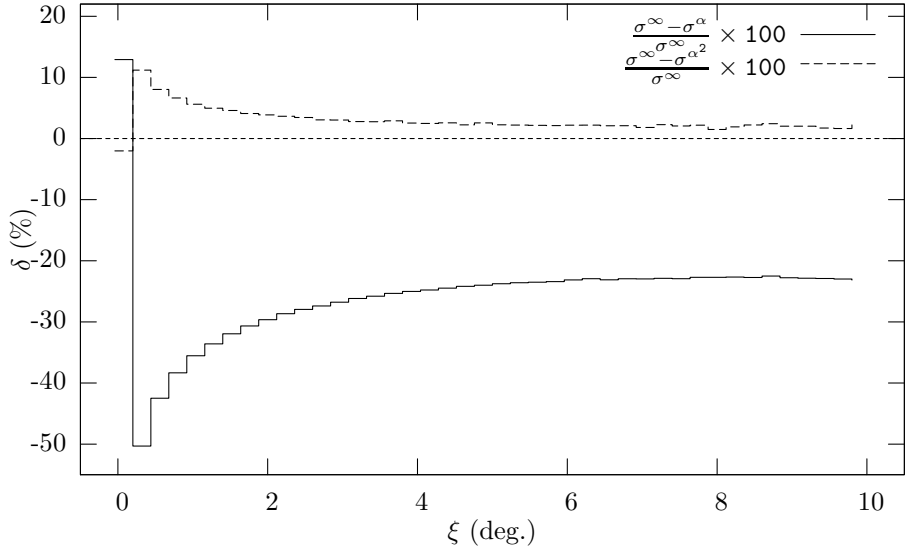}
\caption{Impact of $\alpha^2$ (solid line) and higher order (dotted)
corrections on the acollinearity distribution, in set up (b).}
\label{fig:acolla3}
\end{center}
\end{figure}
This can be demonstrated by comparing the fully resummed cross section
of Eq.~\myref{matchedinfty} with its expansion up to order $\alpha^2$,
which, relying on the results of the next Sections, can be
considered a really good approximation of the complete two-loop
calculation. As in Sect.~\ref{tunedcomparisons}, here we switch off
vacuum polarization effects.

In Fig.~\ref{fig:acolla3},
the effect of the higher order corrections, dominated by $\alpha^3$ 
contributions, is shown in comparison with that of the $\alpha^2$ 
corrections on the acollinearity
distribution for set up (b): as can be seen,
the $\alpha^3$ effect can be as large as
$10\%$ in the phase space region of soft photons emission, corresponding to
small acollinearity angles  with almost back-to-back final-state fermions.

\subsection{Error on two real photons emission cross section}
\label{hardhard}
The cross section  $\sigma^{\alpha^2}_{H,H}$ for the emission of two real
photons in \BABAYAGA\ is approximated by the integral over the phase space
of $F_{1,H}F_{2,H}|{\ourcal M}_{2,LL}|^2d\Phi_2$. By means of the $\tt
ALPHA$~\cite{alpha} algorithm, the two real photon amplitude can be
calculated exactly
and then integrated over the (exact) phase space $d\Phi_2$. In
Tab.~\ref{realhh}, the two real photons cross section as obtained
with the approximation of Eq.~\myref{matchedinfty} and with the exact
matrix element are compared. We choose {$\varepsilon=1\times10^{-4}$}
and we successfully verified that the
difference of the cross sections $\Delta$ is independent of the
$\varepsilon$ choice, as expected. The error $\delta^{err}_{H,H}$
is measured in units of the Born cross section.
\begin{table}[thb]
\begin{center}
\begin{tabular}{|c|c|c|c|c|}
\hline
 set up & $\sigma^{\alpha^2}_{H,H}$ (nb) & $\sigma^{\alpha^2}_{\tt
 ALPHA}$ (nb) & $\Delta$ (nb)  & $\delta^{err}_{H,H}$ (\%)\\
\hline
\hline
 (a) & 2189.37(2) & 2189.4(2) & -0.3(2)   &-4(3)$\times 10^{-3}$\\
 (b) & 229.197(2) & 229.20(2) &  0.00(2)  & 0(4)$\times 10^{-4}$\\
 (c) & 44.0719(5) & 44.072(5) & -0.00(5)  & 0(7)$\times 10^{-3}$\\
 (d) & 4.22839(4) & 4.2286(5) &  0.0002(5)& 4(9)$\times 10^{-3}$\\
\hline
\end{tabular}
\end{center}
\caption{Error on two real photon emission cross section.}
\label{realhh}
\end{table}

In Tab.~\ref{realhh}, we remark the small (and negligible)
error, showing {\it a posteriori} that having corrected the real LL
emissions with the factors $F_{H,i}$ in the master formula gives an 
extremely good approximation of the exact \oatwo\ squared amplitude,
at least within the considered event selection criteria.

\subsection{Light pair corrections}
\label{pairs}
An \oatwo\ contribution which is not included in Eq.~\myref{alpha2content},
even approximately, are the so-called pair corrections. Starting
from the photonic one loop corrected diagrams, the virtual pair
diagrams can be obtained by inserting a fermion loop in the correcting
photon propagator. In order to estimate the size of the virtual pair
corrections (VPC), we use the formulae of
Ref.~\cite{virtualpairs}
(approximated at LL accuracy), for $t$ channel Bhabha. VPC develop
terms of the order of $\alpha^2L^3$ which are then cancelled when also
the real pair emissions (RPC) are included. We also estimate the size of the
real pairs using the soft approximation of 
Ref.~\cite{pairarbuzov,realsoftpairs} and setting the maximum
pair energy $\Delta E = 0.2\times\sqrt{s}$, which is compatible with
the requirements of set up (a)--(d) for the energy of the final-state 
leptons. In
Tab.~\ref{tblpairs}, the virtual and real pair cross section are
reported, and the errors $\delta^{err}_{pairs}$ due to missing them in 
\BABAYAGA\ is measured in units of the Born cross section.
It is worth noting that the relative contribution of light pairs emission is at the
level of a few 0.01\%.
\begin{table}[thb]
\begin{center}
\begin{tabular}{|c|c|c|c|}
\hline
 set up & $\sigma^{\alpha^2}_{VPC}$ (nb) & $\sigma^{\alpha^2}_{RPC}$ (nb) & $\delta^{err}_{pairs}$ (\%)\\
\hline
\hline
 (a) & -4.605(3)  & 3.305(3)  & -0.019\\
 (b) & -0.5698(1) & 0.4375(1) & -0.025\\
 (c) & -0.1385(1) & 0.1154(1) & -0.032\\
 (d) & -0.01542(1)& 0.01320(1)& -0.040\\
\hline
\end{tabular}
\end{center}
\caption{Error due to the virtual and real electron pair corrections.}
\label{tblpairs}
\end{table}

Here we consider only electron pairs.
The contribution of muon pairs and hadronic pairs is expected 
to be one order of magnitude smaller with respect to electron pair 
production. 
This problem was investigated in Ref.~\cite{npbpairs99} for 
the small angle Bhabha scattering at LEP1, where the typical 
momentum transfer is of the order of the energies involved at 
flavour factories. In Ref.~\cite{npbpairs99} the global effect 
of muon and hadronic pairs was safely estimated at the level of 
30\% of the contribution of electronic pairs. 

\subsection{Comparisons with virtual plus soft two-loop calculations}
\label{thun}
In order to establish the error on $\sigma^{\alpha^2}_{SV}$ of
Eq.~\myref{alpha2content}, we compare it with recent calculations
appeared in the literature. As mentioned in the introduction, 
there has been important
progress towards the calculation of the full QED two-loop corrections (NNLO)
to Bhabha scattering. An exhaustive report of the status of the two-loop
QED corrections to Bhabha scattering can be found in Ref.~\cite{bfradcor2005}. 
What is actually available is 
the complete two-loop virtual photonic correction in the approximation
of neglecting 
${\ourcal O}(m^2/Q^2)$, where $Q^2$ stands for one of the Mandelstam 
invariants $s$, $t$ and $u$~\cite{penin}. The real radiation contribution 
is treated in the soft photon approximation and the calculation is 
differential in the electron scattering angle. The results have been 
confirmed by independent calculations~\cite{arbuzovsherbak,boncianiferroglia}. 
In Ref.~\cite{boncianiferroglia} the electron mass terms have been
included but the two-loop box 
diagrams have been neglected. Work is in progress towards the
calculation of massive two-loop box diagrams~\cite{box}. Another
ingredient towards the 
two-loop complete calculation is the virtual fermionic contribution 
$N_f = 1$~\cite{boncianietal} including all finite fermion mass effects. 
In Ref.~\cite{boncianiferroglia} also real pair 
production~\cite{pairarbuzov} and photon emission 
in the soft limit have been introduced 
verifying the cancellation of infrared singularities as well as 
terms proportional to $L^3$ between virtual and real 
corrections. 

We remark that, in order to achieve a NNLO accuracy in Bhabha
scattering, also the exact two real photons corrections and the exact one-loop
corrections to the one real photon emission process must be
included~\cite{realsoftpairs,jadach_vb}.
The complete calculation of the latter 
in particular is still missing in the literature for the large-angle
Bhabha scattering.

The cross section $\sigma^{\alpha^2}_{SV}$ of
Eq.~\myref{alpha2content}
%
can be directly compared with the results of
Ref.~\cite{penin,boncianietal,boncianiferroglia}, 
in order to quantify the size of the missing terms. 
In the following subsections the uncertainties 
inherent to the classes of pure photonic and $N_f = 1$ corrections are 
separately investigated numerically. 

As a first step we write explicitly $\sigma^{\alpha^2}_{SV}$, which is
derived from the first term ($n=0$) of
the infinite sum in Eq.~\myref{matchedinfty}. In order to show the $s$,
$t$ and interference contributions, we define 
\bea
d\sigma^\alpha_{SV}&=&d\sigma^\alpha_{s,SV}
+d\sigma^\alpha_{t,SV}+d\sigma^\alpha_{st,SV}\equiv(E_s+E_t+E_{st})d\sigma_0
\nonumber\\
d\sigma_{0}&=&d\sigma_{s,0}
+d\sigma_{t,0}+d\sigma_{st,0}\equiv(B_s+B_t+B_{st})d\sigma_0
\eea
Truncating 
every factor in Eq.~\myref{matchedinfty}, improved with vacuum
polarization effects as described in Sect.~\ref{vpsubsection}, at ${\ourcal O}(\alpha^2)$ we get
\begin{eqnarray}
\frac{d\sigma_{SV}}{d\sigma_0}
   &\simeq& \left(1 + V + \frac{V^2}{2}\right) \nonumber \\
   &\times& \left[ 1+ (E_s - V B_s) r_s^2 + (E_t - V B_t) r_t^2 
                 +(E_{st} - V B_{st}) r_s r_t \right] \nonumber \\
   &\times& \left( B_s r_s^2 + B_t r_t^2 + B_{st} r_s r_t \right)
\label{beforeexpansion}
\end{eqnarray}
where $V = -(2 \alpha /\pi) I_+ L'$ is the ${\ourcal O}(\alpha)$ term of
the Sudakov form factor, 
$E_i$ and $B_i$ have been defined above and
$r_{s,t}$ are the vacuum 
polarization corrections (including only electron loop to be
consistently compared with the $N_f=1$ results) for the $s$ and $t$ 
channels. If we define $1/(1 -
\Delta\alpha(q^2))\equiv1/(1-\delta_{q^2})$, $r^2_S$, $r^2_t$ and
$r_sr_t$ read:
\begin{eqnarray}
r_s^2 &=& 1 + 2 \delta_s + 3 \delta_s^2 \nonumber \\
r_t^2 &=& 1 + 2 \delta_t + 3 \delta_t^2 \nonumber \\
r_sr_t &=& 1 + \delta_s +\delta_t +\delta_s^2 +\delta_t^2 +\delta_s\delta_t 
\end{eqnarray}
We would like to stress that $\delta_i$ are calculated at one-loop order.

Retaining only terms up to ${\ourcal O}(\alpha^2)$, Eq.~\myref{beforeexpansion} 
reads 
\begin{eqnarray}
\frac{d\sigma_{SV}}{d\sigma_0}
&=& 
1 \nonumber \\
&+& V  + (E_s - V B_s) + (E_t - V B_t) + (E_{st} - V B_{st}) \nonumber \\
&+& 2 (B_s \delta_s + B_t \delta_t) + B_{st} (\delta_s + \delta_t) \nonumber \\
&+& 1/2{V^2} \nonumber \\
&+&  (E_s - V B_s) \delta_s + (E_t - V B_t) \delta_t 
+ (E_{st} - V B_{st}) (\delta_s + \delta_t )\nonumber \\
&+& 3 (B_s \delta_s^2 + B_t \delta_t^2) + B_{st} 
(\delta_s^2 + \delta_t^2 + \delta_s \delta_t) \nonumber \\
&+& V [(E_s - V B_s) + (E_t - V B_t) + (E_{st} - V B_{st}] \nonumber \\
&+& V [2 (B_s\delta_s + B_t\delta_t) + B_{st}(\delta_s +\delta_t)] \nonumber \\
&+& \left[(E_s-V B_s)+(E_t-V B_t)+(E_{st} - V B_{st}\right]\times \nonumber \\
&\times&\left[2(B_s\delta_s+B_t\delta_t)+B_{st} (\delta_s + \delta_t)\right]
\label{expansion}
\end{eqnarray}

The first line of the previous equation is the Born contribution, the
second line is the photonic one loop soft plus virtual correction
(notice that it is equal to $E_s+E_t+E_{st}$ because
$B_s+B_t+B_{st}=1$), the third line is the vacuum polarization
correction at \oa\ and the remaining lines represent $\sigma_{SV}^{\alpha^2}$.

\subsubsection{Two-loop photonic corrections}
\label{cfrpenin}
\begin{figure}[ht]
\begin{center}
\includegraphics{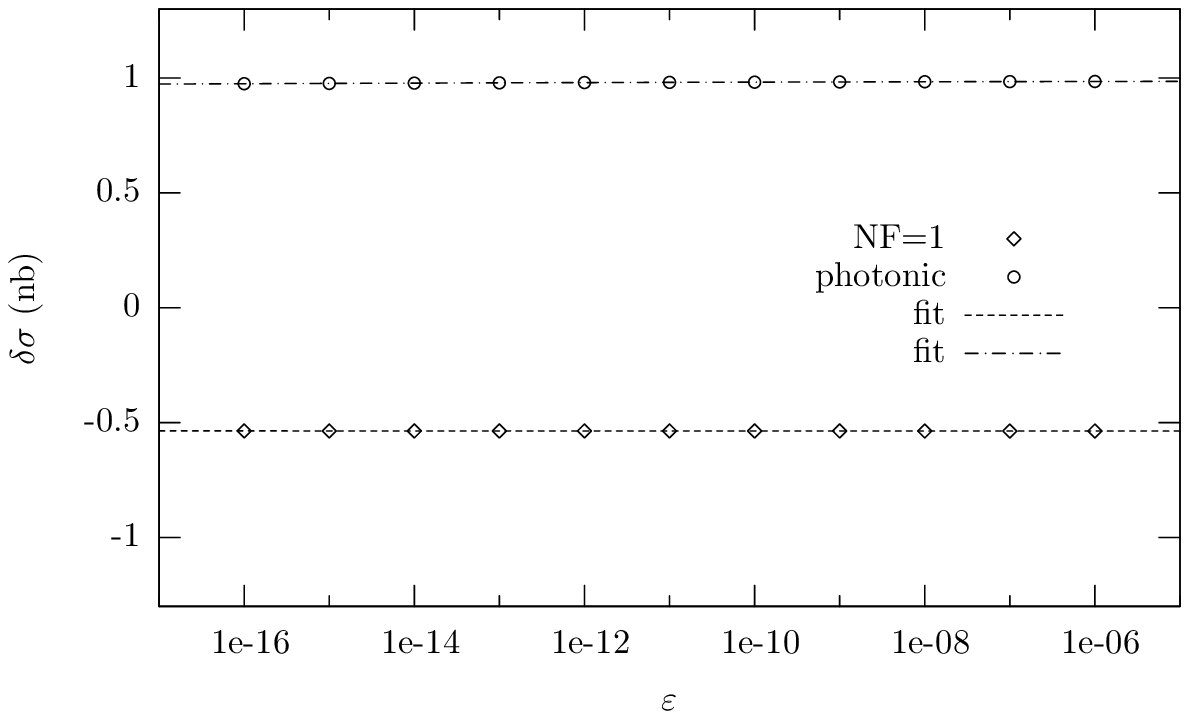}
\caption{Difference between the corrected cross section of 
Eq.~(\ref{expansion}) and the exact two loop photonic and $N_f=1$
corrections, as a function of the 
infrared regulator $\varepsilon$, according to set up (a).}
\label{fig:epsscan}
\end{center}
\end{figure}

After switching off the terms coming from vacuum polarization
contributions $(\delta_s = \delta_t = 0)$, 
the pure ${\ourcal O}(\alpha^2)$ term of the above equation can be
compared with the analytical spectrum 
of Ref.~\cite{penin}. In fact the approximation 
$s$, $t$, $u >> m^2$ is fulfilled for the event selections (a), (b),
(c) and (d) considered in Sect.~\ref{pheno}.
Since all infrared terms are factorized, 
all differences between Eq.~(\ref{expansion}) and the calculation of 
Ref.~\cite{penin} are not expected to be infrared sensitive, apart 
from spurious terms in Eq.~(\ref{expansion}) suppressed by 
coefficients of the order of $m^2/Q^2$~\footnote{In \BABAYAGA, the terms
proportional to $m^2/Q^2$ are only partially accounted for: for
example, the fully massive kinematics is 
always considered, while the non-infrared ${\ourcal O}(m^2/Q^2)$ mass
terms are neglected in the virtual one loop contributions.}.
The infrared behaviour of the difference between Eq.~\myref{expansion} 
and the results of Ref.~\cite{penin}
can be seen in Fig.~\ref{fig:epsscan} (upper curve), 
where the infrared regulator has been 
allowed to scan over a range of ten orders of magnitude for 
set up (a). 
The dashed curve fitting the circles has the following expression:
\begin{eqnarray}
&&\sigma^{\alpha^2,\mbox{\scriptsize \ phot.}}_{SV}
-\sigma^{\alpha^2}_{\mbox{\scriptsize Ref.~\cite{penin}}}
= \left(\frac{\alpha}{\pi}\right)^2 \left(a \frac{m^2}{s}L^2\log^2\varepsilon 
+ b \frac{m^2}{s} L^2 
\log \varepsilon + c L \right) \sigma_0\nonumber \\
&& a = -4.02 \pm 0.01\nonumber\\
&& b = -6.7 \pm 0.4\nonumber \\
&& c   =  +1.75447 \pm 0.00001 
\label{a2lphoteps}
\end{eqnarray}
with $\sigma_0 = 6855.7$~nb and
$L=\log(s/m^2)=15.2$. Equation \myref{a2lphoteps} clearly demonstrates that
in the difference
infrared sensitive contributions survive, but are suppressed by the
factor $m^2/s$. It also shows that, concerning photonic corrections,
the error of \BABAYAGA\ starts at the level of the $\alpha^2L$
corrections, not enhanced by any infrared logarithm~\cite{a2l}.

\begin{figure}[ht]
\begin{center}
\includegraphics{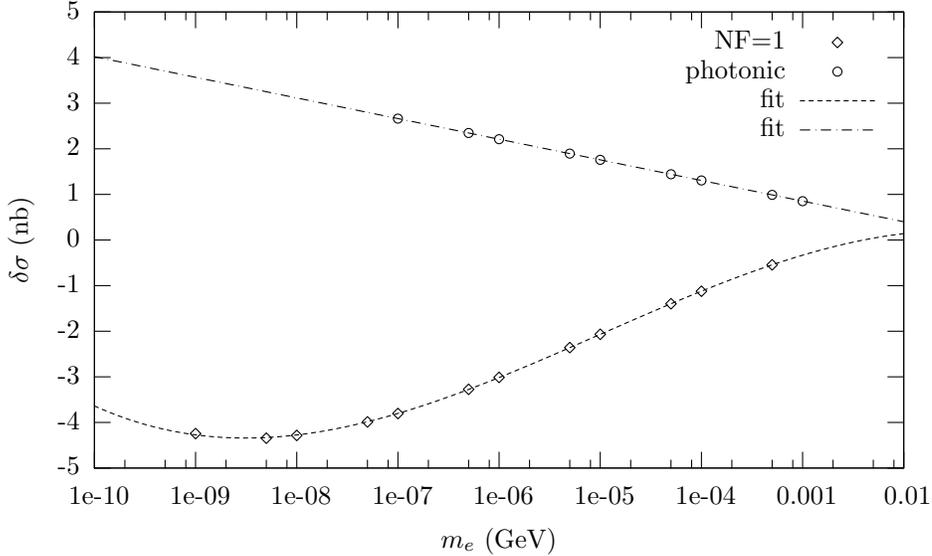}
\caption{Difference between the corrected cross section of 
Eq.~(\ref{expansion}) and the exact two loop photonic and $N_f=1$
corrections, as a function of the electron mass, 
with the infrared regulator $\varepsilon = 1 \times 10^{-5}$, 
according to set up (a).}
\label{fig:mescan}
\end{center}
\end{figure}

In order to numerically check that the term $c L$ in Eq.~(\ref{a2lphoteps}) is 
a true single collinear logarithm, Fig.~\ref{fig:mescan} shows a scan of the 
difference between the QED corrected Bhabha cross section of 
Eq.~(\ref{expansion}) and Ref.~\cite{penin} (upper curve) as a function of the
electron mass, whose values are allowed to span a range of eight
orders of magnitude. The infrared regulator $\varepsilon$ has been
fixed to $1 \times 10^{-5}$. 
The dashed curve fitting the circles has the following expression:
\begin{eqnarray}
&&\sigma^{\alpha^2,\mbox{\scriptsize \ phot.}}_{SV}
-\sigma^{\alpha^2}_{\mbox{\scriptsize Ref.~\cite{penin}}}
= \left(\frac{\alpha}{\pi}\right)^2 \left(a \log\left(
\frac{s}{m^2}\right) + b \pi^2\right) \sigma_0
\nonumber\\
&& a = +2.656  \pm 0.001\nonumber \\
&& b = -1.391 \pm 0.003
\label{a2lphotme}
\end{eqnarray}
showing that the missing two-loop photonic contributions in \BABAYAGA\
are really of the order of $\alpha^2L$. The impact on the integrated
cross section within the realistic set
up of these terms will be shown in Tab.~\ref{table:twoloop}.

\subsubsection{Two-loop $N_f=1$ corrections}
\label{cfrbonciani}
The complete $N_f=1$ virtual calculation of
Ref.~\cite{boncianietal} refers to 
three types of diagrams: 1)~diagrams with a fermionic loop in the tree-level 
photon propagator with a further photonic line within the loop; 
2)~diagrams with a fermionic loop in the photon propagators of the
vertex and box one-loop diagrams; 
3)~diagrams with a fermionic loop in the tree-level diagrams 
with an additional photonic vertex correction. 
In the literature, type 2)~diagrams are referred to as 
virtual pair corrections, while type 3)~diagrams can be thought of as
factorized in vacuum polarization times one-loop photonic
correction contributions.

The part of $\sigma^{\alpha^2}_{SV}$ to be compared with two-loop
$N_f=1$ corrections is extracted 
from Eq.~\myref{expansion} by retaining all terms of photonic and 
vacuum polarization origin. In order to compare with the results of
Ref.~\cite{boncianiferroglia}, we subtract from them
the virtual pair corrections to the
$t$-channel Bhabha scattering already considered in
Sect.~\ref{pairs}~(\cite{virtualpairs}).
Furthermore, the soft real
pair corrections included in Ref.~\cite{boncianiferroglia}
have been switched off, in order to perform a consistent comparison. 

We expect that all infrared sensitive terms have 
the correct coefficient, owing to the factorization of infrared corrections, 
as can be seen in Fig.~\ref{fig:epsscan} (lower curve). The analytical expression of 
the dotted curve fitting the diamonds is given by 
\begin{eqnarray}
&&\sigma^{\alpha^2,\mbox{\scriptsize \ $N_f=1$}}_{SV}
-\left[\sigma^{\alpha^2}_{\mbox{\scriptsize Ref.~\cite{boncianiferroglia}}}-\sigma^{\alpha^2}_{VPC}\right]
=
\left(\frac{\alpha}{\pi}\right)^2 \left(a \frac{m^2}{s} L^2 \log \varepsilon 
+ b L \right)\sigma_0\nonumber \\
&& a = -8.72 \pm 0.07\nonumber \\
&& b = -0.955489 \pm 0.000006
\label{a2lnfeps}
\end{eqnarray}
showing that the two-loop $N_F=1$ infrared structure in \BABAYAGA\ is
under control.

Concerning the dependence from the (logarithm of the) electron mass, apart from
spurious coefficients of order $m^2/Q^2$ and a small residual $\alpha^2L^3$
dependence due to VPC to $s$-channel Bhabha, the terms containing collinear 
logarithms have two sources, namely squared vacuum polarization and interference 
between vacuum polarization and photonic corrections.
Thus the difference between $\sigma^{\alpha^2, \ N_f=1}_{SV}$
and the calculation of Ref.~\cite{boncianiferroglia}
(subtracted of the VPC of Sect.~\ref{pairs}) should start 
from terms of the order of $\alpha^2L^2$. In fact, the dotted curve fitting
the diamonds of Fig.~\ref{fig:mescan} has the following form:
\begin{eqnarray}
&&\sigma^{\alpha^2,\mbox{\scriptsize \ $N_f=1$}}_{SV}
-\left[\sigma^{\alpha^2}_{\mbox{\scriptsize Ref.~\cite{boncianiferroglia}}}-\sigma^{\alpha^2}_{VPC}\right]\nonumber=\\
&&=\left(\frac{\alpha}{\pi}\right)^2 \left(a \log^3 \left( \frac{s}{m^2}\right)
 + b \log^2 \left( \frac{s}{m^2} \right) + c \log \left( \frac{s}{m^2} \right)
 + d \pi^2 \right)\sigma_0 
\nonumber\\
&& a = +0.00728 \pm 0.00007\nonumber \\
&& b = -0.504 \pm 0.005\nonumber \\
&& c = +5.9 \pm 0.1\nonumber \\
&& d = -1.36 \pm 0.08
\label{a2lnfme}
\end{eqnarray}
In the next Section the size of the error induced by missing exact
photonic and $N_f=1$
corrections in $\sigma^{\alpha^2}_{SV}$ is discussed.

\subsection{Error on purely virtual plus soft two-loop cross section}
The numerical effect on the cross section, for the four different 
set up considered in this study, is shown 
in Tab.~\ref{table:twoloop}, with the infrared separator
$\varepsilon$ set to $10^{-5}$.
In Tab.~\ref{table:twoloop}, the $\Delta$ are the difference of the
exact and the \BABAYAGA\ $\alpha^2$ virtual plus soft cross section
and $\delta^{err}$ are their size in unit of the Born cross section.
\begin{table}[thb]
\begin{center}
\begin{tabular}{|c|c|c|c|c|c|}
\hline
 set up & $\Delta_{phot}$~(nb) &  $\Delta_{N_f=1}$~(nb) & 
$\delta^{err}_{phot}$ (\%)& $\delta^{err}_{N_f=1}$ (\%) & $\delta^{err}_{phot+N_f=1}$ (\%)\\
\hline
\hline
(a)  & 0.9855 &  -0.537(2)   & $0.014$ & $-0.0078$ & $0.0062$\\
(b)  & 0.1218  & -0.0442(3)  & $0.023$ & $-0.0083$ & $0.0147$\\
(c)  & 0.0149 &  -0.01436(4) & $0.021$ & $-0.020$  & $0.001$\\
(d)  & 0.0016 &  -0.00099(1) & $0.029$ & $-0.018$  & $0.011$\\
\hline
\end{tabular}
\end{center}
\caption{Cross section difference ($\Delta_{phot, N_f=1}$) as obtained with
$\sigma^{\alpha^2}_{SV}$ and the exact two loop corrections, for purely
photonic contributions and $N_f=1$. $\delta_{phot,N_f=1}$ are the
differences of the cross sections in units of the Born one.}
\label{table:twoloop}
\end{table}
It is worth noticing the accidental opposite sign between $\delta_{phot}$ 
and $\delta_{N_f=1}$ for all event selections considered, allowing to 
reduce the error of \BABAYAGA\ predictions when summing up 
everything together.
%
\subsection{Error on the virtual plus soft corrections to one photon
real emission}
\label{virtualbrems}
The virtual plus soft cross section with one hard photon
$\sigma^{\alpha^2}_{SV,H}$ of Eq.~\myref{alpha2content} is obtained
from the \oatwo\ content of the term $n=1$ in Eq.~\myref{matchedinfty}. 
Its exact expression for large angle Bhabha scattering is not known
in the literature, but it was calculated for the small angle
Bhabha process~\cite{jadach_vb} or for the
$s$-channel~\cite{vbschannel} at large angles.
Relying on the LEP experience and on the purely virtual plus soft
results described in Sects.~\ref{cfrpenin} and \ref{cfrbonciani}, the
error of the \BABAYAGA\ approximation is at the level of the infrared-safe
$\alpha^2 L$ terms, the size of which can be safely taken to be
smaller than the $0.05\%$ for all the set up.

\subsection{Non-perturbative error induced by hadronic contribution to
 vacuum polarization}
\label{vperror}
By means of the analysis presented in the previous Sections, we can
deduce that the missing \oatwo\ (perturbative) 
contributions in \BABAYAGA\ do not exceed the 0.1\%.

Nevertheless, besides the missing \oatwo\ contributions considered before,
another source of theoretical error, coming from the hadronic
contribution to the vacuum polarization, has to be carefully
considered. This uncertainty has an intrinsically non-perturbative origin.
Since the routine $\tt HADR5N$, by means of
which we calculate $\Delta\alpha^{(5)}_{hadr}(q^2)$, returns also an error
$\delta_{hadr}$ on its value, we estimate the induced error
by computing the cross section with
$\Delta\alpha^{(5)}_{hadr}(q^2)\pm\delta_{hadr}$ and taking the
difference as the theoretical uncertainty due to the hadronic
contribution to vacuum
polarization. In Tab.~\ref{vpunctbl}, the uncertainties $\Delta$ are
calculated for the event selection criteria (a)-(d) on the Born and
the matched cross section. The quantities
$\delta^{err}_{VP}$ are the errors in units of the Born cross
section. 
Differently from previous investigations of vacuum polarization uncertainty
in small-angle Bhabha scattering at LEP~\cite{yrbhabha-95}, we consider here
only the error induced by the parameterization of hadronic loops of
Refs.~\cite{jeg-2003,hadr5}, because of the absence of other results
able to keep under control appropriately the contribution of
hadronic resonances for time-like momenta circulating in the
photon self-energy.

\begin{table}[thb]
\begin{center}
\begin{tabular}{|c|c|c|c|c|}
\hline
 set up & $\Delta_{Born}$ (nb) & $\Delta_{full}$ (nb) & 
$\delta^{err}_{VP,Born}$ (\%)  & $\delta^{err}_{VP,full}$ (\%)\\
\hline
\hline
 (a) & -0.48    & -0.50   & -0.007  & -0.007\\
 (b) & -0.00070 & -0.0014 &  0.     &  0.   \\
 (c) &  0.017   &  0.014  &  0.024  &  0.020\\
 (d) &  0.0033  &  0.0024 &  0.060  &  0.044\\
\hline
\end{tabular}
\end{center}
\caption{Error induced by the uncertainty on $\Delta\alpha^{(5)}_{hadr}(q^2)$.}
\label{vpunctbl}
\end{table}
It must be remarked that the error on $\Delta\alpha^{(5)}_{hadr}(q^2)$
for $q^2>0$ strongly varies and increases passing through hadronic
resonances. This yields a worsening of the theoretical accuracy of Bhabha cross
section, in particular close to the $c\bar{c}$ states in the region
between 3 and 4.5~GeV
which have a non-negligible branching ratio into electrons. 
Relying on the output of the $\tt HADR5N$ routine (which here prints a
warning about its reliability), we verified that the error on the
Bhabha cross section induced by hadronic loop may reach the 0.5\%
in this region and it is therefore the limiting factor for the
theoretical accuracy.

\subsection{Summary of the theoretical errors}
\label{summaryoferrors}
The size of the $\alpha^2$ contributions missing in \BABAYAGA\ and the
error induced by the hadronic contribution to the vacuum polarization,
within the realistic event selection criteria considered in this
paper, are summarized in Tab.~\ref{errorsummarytable}.
\begin{table}[thb]
\begin{center}
\begin{tabular}{|c|c|c|c|c|}
\hline
$|\delta^{err}|$ (\%) & (a) & (b) & (c) & (d)\\
\hline
\hline
 $|\delta^{err}_{VP}|$        & 0.01 & 0.00 & 0.02 & 0.04\\
 $|\delta^{err}_{pairs}|$     & 0.02 & 0.03 & 0.03 & 0.04\\
 $|\delta^{err}_{H,H}|$       & 0.00 & 0.00 & 0.00 & 0.00\\
 $|\delta^{err}_{phot+N_f=1}|$& 0.01 & 0.01 & 0.00 & 0.01\\
 $|\delta^{err}_{SV,H}|$      & 0.05 & 0.05 & 0.05 & 0.05\\
\hline
 $|\delta^{err}_{total}|$     & 0.09 & 0.09 & 0.10 & 0.14\\
\hline
\end{tabular}
\end{center}
\caption{Summary of different source of theoretical error in \BABAYAGA.}
\label{errorsummarytable}
\end{table}

The size of the virtual and real pair corrections and of the
missing virtual corrections to the real photon emission process is
only guessed with a safe estimate of their impact. We also remark that
the somehow large error coming from vacuum polarization uncertainty in
set up (c) and (d) is due to the fact that we consider energies around
the $\Upsilon$ resonance and the data-driven routine $\tt HADR5N$
produces here larger errors.

From Tab.~\ref{errorsummarytable}, by summing up the absolute value of all
the uncertainties, we can deduce that a safe estimate
of the total theoretical accuracy of the new version of \BABAYAGA\ (based on
the master formula~\myref{matchedinfty} and the $\tt HADR5N$ routine
for the hadronic contribution to the vacuum polarization) 
for the calculation of the Bhabha cross section is 0.1\% at $\Phi$ 
factories and below 0.2\% at the $B$ factories.

We remark that concerning the perturbative \oatwo\ error, our
formulation is able to reach the $0.1\%$ accuracy in all the
considered set up and that, as previously emphasized, the accuracy can be 
worsened to the 0.5\% level in proximity of the $J/\Psi$ 
resonances because of the error on the
non-perturbative hadronic contribution to the vacuum polarization.

%% file: conclusions.tex
\section{Conclusions}
\label{conclusion}
Bhabha scattering is a crucial process for precise luminosity monitoring of
$e^+ e^-$ colliders. Ongoing and future experiments at $e^+ e^-$ accelerators
operating in the region of hadronic resonances, such as
$\Phi$, $\tau$-charm
and $B$ factories, require a precise luminosity knowledge,
especially in view of improved measurements of the hadronic cross section in 
the energy region going from the pion pair production threshold up to 12~GeV.

With this motivation in mind, we have presented in this paper a 
high-precision calculation of the Bhabha
process in QED, with a precision target of the order of 0.1\%. This
theoretical accuracy is achieved by the
matching of exact NLO corrections with higher-order leading
logarithmic contributions, which
are kept under control through all orders of $\alpha$ by means of a
QED Parton Shower approach.
The matching of matrix element corrections with Parton Showers is a
topic of QCD calculations and recently different successful solutions
have been proposed, opening the way towards
precision calculations of high-energy QCD processes
\cite{ns-2006}. Our matching algorithm
represents, to the best of our knowledge, the first example of
such an application in QED.
The theoretical precision of the approach has been demonstrated
through detailed comparisons with
the predictions of precise independent generators and, noticeably,
with the results of recent calculations
of two-loop corrections to Bhabha process, which are the frontier of
modern perturbative QED.
Other components of the theoretical luminosity
error, coming from hadronic vacuum polarization and light pair
emission, has been 
carefully estimated as well, in order to arrive at a sound and robust
total error budget.
In particular, the perturbative contribution to the error coming from
missing \oatwo\ corrections has been shown to be under control at the
0.1\% level for all the experimental selection criteria here
considered, realistic for data analysis at flavour
factories. Furthermore, the non-perturbative error due to the
uncertainty on the hadronic contribution to vacuum polarization, which
critically depends on the centre-of-mass energy, turns out to be
below the 0.05\% level at $\Phi$ and $B$ factories, whereas can
increase up to the 0.5\% in proximity of the $c\bar{c}$ bound
states. The reduction of such an uncertainty, if needed, requires
progress both on the theory and the experimental side.

The theoretical precision for luminosity monitoring through
large-angle Bhabha scattering at flavour factories is now comparable
with the accuracy reached at the time of the LEP workshop working
groups 95/96 for the theoretical error underlying the 
high-precision small-angle Bhabha measurement at 
LEP~\cite{yrbhabha-95,aetal-96}.

The new formulation has been implemented in an improved version of the 
event generator \BABAYAGA~\cite{sitobabayaga},
which is available for high-precision luminosity measurements at flavour 
factories. For example, this will allow to reduce the theoretical
error in the luminosity
measurement at the $\Phi$ factory DA$\Phi$NE from the present 0.5\% to
0.1\%, thus roughly halving the total luminosity error quoted by KLOE
collaboration at DA$\Phi$NE and, more generally, paving the road to
more precise measurements of the Bhabha process at other 
$e^+ e^-$ collider, such as BEPC, CESR, KEK-B, PEP-II and VEPP-2M.

As far as possible future applications of the approach here presented
are concerned, it would certainly be of interest for future data
analysis extending the present phenomenological analysis
to the production processes of muon and photon pairs in $e^+e^-$
annihilation, as well as to their exclusive signatures such as
$e^+e^-\to\mu^+\mu^-\gamma$, because also these
QED reactions are employed at flavour factories for precise luminosity
studies~\cite{sighad2003}.  Other
interesting perspectives would be the applying the matching
procedure
to the Standard Model processes presently under consideration
for luminosity monitoring of a future $e^+ e^-$
collider at the TeV scale and extending the matching to incorporate
also the full two-loop corrections.

These developments are by now under consideration.

\section*{Acknowledgments}
We are grateful to M. Moretti for the help in the use of the $\tt ALPHA$
code and for many useful discussions and to R.~Bonciani, A.~Ferroglia
and A.~Vicini for having carefully 
read the preliminary manuscript. We are
indebted with R.~Bonciani, A.~Ferroglia
and A.~Penin
for providing us the numerical program implementing the two-loop calculations.
We also wish to thank
A.~Denig, S.~Eidelman, F.~Jegerlehner, F.~Nguyen and G.~Venanzoni
for many informative discussions and continuous interest in our
work.

%% file: appendix.tex
\section{Technical details}
\label{appendice}
Here we discuss some technical details about Eq.~\myref{matchedinfty}
and its integration, namely the independence from $\varepsilon$,
the mapping of the momenta needed for $n\ge 2$ and the importance sampling of
the final-state collinear singularities.

\subsection{Independence of the master formula from the infrared separator}
Considering Eq.~\myref{generalLL}, its independence from $\varepsilon$
can be demonstrated analytically if we neglect the variation of
$|{\ourcal M}_0|^2$ from the center of mass energy. In this case,
integrating over all the phase space, we get
\be
\sigma \simeq\exp(-\frac{\alpha}{2\pi}I_+L^\prime)\sum_{n=0}^\infty
\left(\frac{\alpha}{2\pi}\right)^n L^{\prime n} I_+^n \sigma_0 = \sigma_0
\ee
Considering now Eq.~\myref{matchedinfty}, we can write the integral of
the radiation factor for each emitted photon as
$\frac{\alpha}{2\pi}L^\prime (I_+ + c)$, where $c$ is a constant not
singular as $\varepsilon$ goes to zero and coming from the correction
factor $F_H$. Thus, the total cross section can be written as
\bea
\sigma &\simeq& F_{SV}\nonumber
\exp(-\frac{\alpha}{2\pi}I_+L^\prime)\sum_{n=0}^\infty
\left(\frac{\alpha}{2\pi}\right)^n L^{\prime n} \left(I_++c\right)^n
\sigma_0\\
&=& F_{SV}\exp(\frac{\alpha}{2\pi}c) \sigma_0
\eea

which does not depend on $\varepsilon$. It is worth noticing that
having corrected each hard photon emission in Eq.~\myref{matchedinfty}
with the factors $F_{H,i}$ is crucial for the good infrared behaviour
of the integrated cross section. In Fig.~\ref{fig:fig1} of Sect.~\ref{pheno} the numerical
independence from the infrared separator has been shown.

\subsection{Mapping of the momenta for $n\ge 2$}
In the master formula~\myref{matchedinfty}, the sum over all possible
photon multiplicities is required, but the building blocks of the
matched cross section (i.e. the squared matrix elements) are strictly
defined only for $0$ or $1$ photon
in the final state. It is therefore mandatory to devise an algorithm to map a
$n$ photons momenta configuration to a $0$ or $1$ photon
configuration in order to consistently calculate the squared
amplitudes $|{\ourcal M}_0|^2$ and $|{\ourcal M}_1|^2$. The mapping algorithm is
not unique, but the choice among different mappings gives a higher
order effect which has been verified to be negligible.

When a photon multiplicity $n$ is selected,
the
first step to define a zero-photon kinematics configuration is to
associate $n_I$ photons as emitted by the initial
state and $n_F$ by the final state. The association is done according
to an algorithm finding whether the photon is nearer to an
initial-state or
final-state fermion, in terms of its angle with respect to the fermion. If
$K_I=\sum_{j=1}^{n_I} k_j$ and $q=p_1+p_2 - K_I$, two initial-state
mapped momenta are defined such as $(p_{1,M}+p_{2,M})^2 = q^2$ and
$p_{1,M}^2= p_{2,M}^2 = m^2$. 
The mapped final-state momenta
are defined such as $(p_{3,M}+p_{4,M})^2 = (p_{1,M}+p_{2,M})^2$,
$p_{3,M}^2= p_{4,M}^2 = m^2$ and $p_{3,M}$ is directed along the
direction of $p_3$ boosted in the frame where $p_3+p_4$ is at rest.
The mapped momenta, which satisfy by
construction momentum conservation and on-shell relations, are used to
calculate the Lorentz-invariant Born squared amplitude needed to
compute $|{\ourcal M}_{n,LL}|^2$ in Eq.~\myref{matchedinfty}.

In order to calculate the correction factors $F_{H,i}$, a mapping to
$1$ photon configuration is needed.
Suppose we are calculating $F_{H,l}$ for the $l$-th
photon: in order to get the mapped momenta, we do as if only
this photon is present.
We keep $p_{1,M}=p_1$,
$p_{2,M}=p_2$ and $k_{l,M}=k_l$, we boost $p_3$ and $p_4$ where
$p_3+p_4$ is at rest, we calculate $p_{3,M}$ and $p_{4,M}$ such as
$(p_{3,M}+p_{4,M})^2 = (p_{1,M}+p_{2,M}-k_{l,M})^2$,
$p_{3,M}^2= p_{4,M}^2 = m^2$ and $p_{3,M}$ is directed along $p_3$ in
this frame and finally we boost $p_{3,M}$ and $p_{4,M}$ back in the
original frame. 
We obtain a set of mapped momenta satisfying momentum conservation and
on-shell relations which can be used to calculate $|{\ourcal M}_1|^2$ and
$|{\ourcal M}_{1,LL}|^2$ and the correction factors $F_{H,i}$.

\subsection{Importance sampling of final-state collinear singularities}
The integral over the phase space $d\Phi_n$ is done by means of Monte
Carlo techniques. For $n$ photons in the final state, the number of independent
integration variables is $3n+2$, and we choose them to be the azimuthal
and the polar angles of one of the final-state fermions and the
energies and the angles of the $n$ emitted photons. With this choice,
$d\tilde{\Phi}_n\equiv d\Phi_n\times flux$ can be written as
\be
d\tilde{\Phi}_n=\frac{k^0_1\cdots  k^0_n}{(2\pi)^{3n+2}2^{n+2}}
\frac{|\vec{p_3}|}{p^0_4+p^0_3\left(1+
\frac{\vec{K}\cdot\vec{p_3}}{|\vec{p_3}|^2}\right)}
dk^0_1\cdots dk^0_n d\Omega_{\gamma 1}\cdots d\Omega_{\gamma n} d\Omega_3
\label{dfnesplicito}
\ee
where $K\equiv\sum_{i=1}^nk_i$ and the final-state momenta are easily
calculated with the independent variables by requiring momentum
conservation and on-shell relations. In Eq.~\myref{dfnesplicito}, the
angles of the final-state electron have been chosen, but the positron
could have been equivalently chosen.

The generation of the independent variables is done according to the
peaking structure of the function to be integrated. Here we discuss
in some detail the sampling of the final-state collinear
singularities. The peaks of
the differential cross section can be read from
$|{\ourcal  M}_{n,LL}|^2d\Phi_n$ (the correction factors $F_{H,i}$ tend to $1$
in the singular regions), namely
\be
|{\ourcal M}_{n,LL}|^2d\Phi_n\propto dk^0_1\cdots dk^0_n 
d\Omega_{\gamma 1}\cdots d\Omega_{\gamma n} d\Omega_3\frac1{k^0_1\cdots k^0_n}
I(k_1)\cdots I(k_n)
\label{singularbehaviour}
\ee
In the previous equation the infrared and collinear singularities for each
emitted photon are evident. If infrared singularities do not
present problems (photon energies have to be sampled as $1/k^0$),
collinear peaks, appearing in $\prod_{i=1}^n I(k_i)$, have to be
carefully treated when $n\ge 2$. 

Firstly, $\prod_{i=1}^n I(k_i)$ is flattened with the function 
$\prod_{i=1}^n \tilde{I}(k_i)$, where $\tilde{I}(k)\equiv\sum_{i=1}^4
\frac{1}{p_i\cdot k}$ has the same leading singularities of the
original function. We then write
\bea
\prod_{i=1}^n I(k_i)&=&
\prod_{i=1}^n \frac{I(k_i)}{\tilde{I}(k_i)}
\tilde{I}(k_i) =\nonumber\\ 
&=&\prod_{i=1}^n \frac{I(k_i)}{\tilde{I}(k_i)}
\prod_{j=1}^n\left(
\frac{1}{p_1\cdot k_j}+
\frac{1}{p_2\cdot k_j}+
\frac{1}{p_3\cdot k_j}+
\frac{1}{p_4\cdot k_j}
\right)
\eea
We have to sample the second product in the right hand side of the
above relation, which can be expanded as a sum of terms. The sum is done
via Monte Carlo by choosing randomly for each of the photons a fermion to
generate its angle according to $1/{p\cdot k}$. If $n_i$ is the number
of photons ``attached'' to the fermion $i$, a single term of the sum
takes the form
\be
\prod_{i=1}^{n_1}\frac1{p_1\cdot k_i}
\prod_{j=1}^{n_2}\frac1{p_2\cdot k_j}
\prod_{l=1}^{n_3}\frac1{p_3\cdot k_l}
\prod_{m=1}^{n_4}\frac1{p_4\cdot k_m}
\label{prodterm}
\ee
The sampling of the angular structure of the previous formula is not
trivial. The initial-state sampling (i.e. $1/p_{1,2}\cdot k$) is easy, 
being the momenta $p_{1,2}$ fixed, while to sample final-state
singularities the following problem arises: considering for example
the case $n=2$ with $n_3=1$ and $n_4=1$ and that we choose the angles
of $p_3$ as independent variables for the phase space integral,
Eq.~\myref{prodterm} says we have to
generate the $k_1$ angles along the $p_3$ direction
according to
$\propto 1/p_{3}\cdot k_1$
and the $k_2$ angles along $p_4$.
The point is that the direction of $p_4$ can not be defined until all the
other momenta are generated. We by-pass the problem by using a
multi-channel approach. In the case we are considering, we write
\bea
\frac1{p_3\cdot k_1}\frac1{p_4\cdot k_2} &=& 
\frac1{p_3\cdot k_1}\frac1{p_4\cdot k_2}
\frac{\frac1{p_3\cdot k_1}}{\frac1{p_3\cdot k_1}+\frac1{p_4\cdot k_2}}+\nonumber\\
&&+\frac1{p_3\cdot k_1}\frac1{p_4\cdot k_2}
\frac{\frac1{p_4\cdot k_2}}{\frac1{p_3\cdot k_1}+\frac1{p_4\cdot k_2}}
\label{n3n41}
\eea
The above sum is done again by Monte Carlo, integrating the phase
space by using as independent
variables the angles of the final-state electron ($p_3$) in the first
term of the sum and the angles of the final-state positron ($p_4$)
in the second one. In the first term we generate $k_1$ along $p_3$ and
$k_2$ along a direction which is not exactly that of $p_4$, being the
$1/p_4\cdot k_2$ peak flattened in any case by the denominator
$1/p_3\cdot k_1 + 1/p_4\cdot k_2$. In the second term the role of
$p_3$ and $p_4$ are exchanged.

The generalization of Eq.~\myref{n3n41} to the case $n_3$ and $n_4\neq 1$ reads
\bea
\prod_{i=1}^{n_3}\frac1{p_3\cdot k_i}\prod_{j=1}^{n_4}\frac1{p_4\cdot k_j}&=& 
\prod_{i=1}^{n_3}\frac1{p_3\cdot k_i}\prod_{j=1}^{n_4}\frac1{p_4\cdot k_j}
~\frac{\prod_{i=1}^{n_3}\frac1{p_3\cdot k_i}}
{\prod_{i=1}^{n_3}\frac1{p_3\cdot k_i}+\prod_{j=1}^{n_4}\frac1{p_4\cdot k_j}}
+\nonumber\\
&+&\prod_{i=1}^{n_3}\frac1{p_3\cdot k_i}\prod_{j=1}^{n_4}\frac1{p_4\cdot k_j}
~\frac{\prod_{j=1}^{n_4}\frac1{p_4\cdot k_j}}
{\prod_{i=1}^{n_3}\frac1{p_3\cdot k_i}+\prod_{j=1}^{n_4}\frac1{p_4\cdot k_j}}
\label{n3n4}
\eea
which suggests a similar generation of the independent variables.